\documentclass[a4paper,11pt]{article}
\usepackage[titletoc]{appendix}

\usepackage[utf8]{inputenc}
\usepackage{amsfonts}
\usepackage{amsmath}
\usepackage{amssymb}
\usepackage{amsthm}
\usepackage{url}
\usepackage[english]{babel}
\usepackage{verbatim}
\usepackage{graphicx}
\usepackage{enumerate}
\usepackage{bbm}
\usepackage[retainorgcmds]{IEEEtrantools}

\usepackage{hyperref}

\usepackage{tikz}
\usetikzlibrary{shapes,arrows,chains}

\makeatletter
\def\namedlabel#1#2{\begingroup
    #2%
    \def\@currentlabel{#2}%
    \phantomsection\label{#1}\endgroup
}
\makeatother

\numberwithin{equation}{section}

\def\e{{\epsilon}}

\def\esssup{\hbox{ess sup}}
\def\half{\frac{1}{2}}
\def\mp{\mu_+}
\def\sp{\sigma_+}
\def\sm{\sigma_-}
\def\mm{{\m_-}}
\def\mo{{\m_0}}
\def\so{{\sigma_0}}
\def\sm{{\sigma_-}}

\def\erf{\text{erf}}

\def\hpsi{\hat\psi}
\def\hphi{\hat\phi}

\def\psig{\psi^+_\gamma}
\def\phig{\phi^+_\gamma}

\def\m{\mu}
\def\sig{\sigma}

\def\mL{{\mathbb{L}}}

\def\I{\mathcal{I}}
\def\T{{\mathcal T}}
\def\G{{\mathcal G}}

\def\L{{\mathcal L}}
\def\F{{\mathcal F}}
\def\bF{{\mathbb F}}
\def\P{{\mathit P}}

\def\R{{\mathbb R}}

\def\B{{\mathcal B}}
\def\E{{\mathbb E}}

\def\D{{\mathcal D}}
\def\tV{{\tilde {\V}}}
\def\ts{{\tilde s}}
\def\t{\theta}

\def\l{{\lambda_0}}
\def\la{\lambda}
\def\m{\mu}
\def\stop{\varsigma}
\def\a{\alpha}
\def\b{\beta}
\def\t{\tau}

\def\1{{\mathbbm{1}}}
\def\e{\epsilon}
\def\gg{\mathbb{A}}
\def\ll{\mathbb{B}}
\def\r{{\rho}}
\def\lef{{\nu}}
\def\q{Q}
\def\H{K}

\def\tg{{\tilde g}}
\def\tu{{\tilde u}}
\def\tD{{\tilde D}}
\def\hD{{\hat D}}
\def\tC{{\tilde C}}
\def\hC{{\hat C}}

\def\V{{\mathbf V}}
\def\func{{v}}
\def\W{{\mathbf W}}
\def\low{{\delta}}

\def\tW{{\tilde \W}}

\theoremstyle{plain}
\newtheorem{thm}{Theorem}[section]
\newtheorem{lemma}[thm]{Lemma}

\newtheorem{cor}[thm]{Corollary}
\newtheorem{example}[thm]{Example}
\theoremstyle{remark}
\newtheorem{remark}[thm]{Remark}
\theoremstyle{definition}
\newtheorem{defn}[thm]{Definition}
\theoremstyle{remark}
\newtheorem{assumption}[thm]{Assumption}

\title{The Support and Resistance Line Method: \\  An Analysis via Optimal Stopping}
\usepackage{authblk}
\author{Vicky Henderson \qquad Saul Jacka \qquad Ruiqi Liu \qquad Jun Maeda\footnote{
Department of Statistics, University of Warwick, Coventry, CV4 7AL. UK. \\
Email: \\
vicky.henderson@warwick.ac.uk,
s.d.jacka@warwick.ac.uk, Ruiqi.Liu1994@outlook.com,
jun.maeda@warwick.ac.uk }
\thanks{We would like to thank participants at the Leeds Winter School on Theory and Practice of Optimal Stopping and Free Boundary Problems (13-17 January 2020) and at the Applied Probability conference in honour of Michael Katehakis and Isaac Sonin (21-22 April 2023) for helpful comments. We especially thank the two anonymous referees and Associate Editor for their substantial input. 
}
}

\begin{document}
\maketitle
\tikzstyle{process} = [rectangle, minimum width=3cm, minimum height=0.5cm, text width=8em,text centered, thick,draw=black]
\tikzstyle{decision} = [diamond, minimum width=2cm, minimum height=0.5cm,text width=9em, text centered, aspect=2.5,thick,draw=black]
\tikzstyle{arrow} = [thick,->,>=stealth]

\begin{abstract}
We study a mathematical model motivated by the support/resistance line method in technical analysis where the underlying stock price transitions between three states of nature in a path-dependent manner. For optimal stopping problems with respect to a general class of reward functions and dynamics, using probabilistic methods, we show that the value function is $C^1$  
and solves a general free boundary problem. Moreover, for a range of utilities, we prove that the best time to buy and sell the stock are obtained by solving free boundary problems corresponding to two linked optimal stopping problems. We use this to compute optimal trading strategies for several types of dynamics and varying degrees of relative risk aversion. 
\end{abstract}

\vspace{0.3cm}
\noindent {\bf{Keywords:}}
Optimal stopping, technical analysis, resistance level, support line. 

\noindent {\bf{JEL:}} G11; C61; D53; D91. 
\newline {\bf MSC:} 60G40; 91B24; 91G80.
 
\section{Introduction}\label{section: Introduction}

Technical analysis (TA) is a method to identify trading opportunities by analysing historical market data and price patterns. Traders believe that by observing key market indicators and charts they can predict future price movement, and construct profitable trading strategies. TA is extremely popular among investors. In a survey of 678 fund managers Menkhoff \cite{Men} found that 86\% of fund managers rely on TA as one of their investment tools. Hoffmann and Shefrin \cite{Hof} analyze survey responses from individual investors and report that 32\% use TA.

Through the development of TA, numerous trading rules have been introduced. For example, traders may generate buy/sell signals by comparisons of short and long-term moving-averages; from breakthroughs of market support and resistance levels; from so-called Bollinger bands, and from directional indicators. 
Despite the richness of technical trading strategies, many of them have been criticised for being subjective and lacking mathematical justification. Furthermore, TA is also contentious due to the perception of conflict between its claimed predictive power and the Efficient Markets Hypothesis (Park and Irwin \cite{Par}). 

In this paper we study the prescriptions of the support/resistance line method. Under this method, traders usually buy (sell) an asset if its price goes below (above) a support (resistance) level, or simply, ``buy at low'' (BL) and ``sell at high'' (SH), the so-called {\em standard trading rule}. The support (resistance) line is viewed as a local minimum (maximum) of the asset price over a period of time. However, when the price goes \emph{substantially} below (above) the support (resistance) line, it is said to have {\em broken-through} the line and it is widely accepted that the support (resistance) line will become the new resistance (support) line because of the negative (positive) outlook for the asset resulting from such a price movement. Trading using the insights of the support/resistance line method will result in highly path-dependent strategies which depend upon the past local maxima and minima of prices. 

We propose a rich model set-up which is inspired by the support/resistance line method, but employs standard utility functions, 
and we explore the extent to which it makes trading prescriptions in line with the above description. 
The stock price process may be described as follows. We assume there are three regimes for the stock price process, termed the \textit{positive}, \textit{zero} and \textit{negative} regime respectively. The dynamics of the stock price process are dependent on its current regime. We further assume that there is a fixed price level located in some known interval $[L,H]$, and this price level is the support line if the stock is in the positive regime and the resistance line if it is in the negative regime. The regime changes from the negative (positive) to the positive (negative) regime if the stock price crosses $H$($L$) from below (above) and decays from the positive regime to the zero regime at constant rate $\l$. So, when a transition between positive and negative regimes occurs, there is a reversal of the role of the resistance and support level in line with what traders would expect. The exponential decay from the positive to the zero regime reflects the fact that investors would not expect the (advantageous) positive dynamics to endure indefinitely. Note that the stock price process can be in any regime on the interval $(L,H)$, which provides the flexibility to move around the support/resistance line without changing regimes. 
In addition, to incorporate resistance to regime change, we assume partial reflection at the support/resistance line. This is to model the view of TA traders who believe prices reflect with some probability at resistance and support levels. 

Under our modelling framework, we provide very general results for a wide class of dynamics and utility functions. We will rigorously study its mathematical properties with a broad range of reward functions with the aid of probabilistic arguments. We will show, under mild assumptions on reward functions and dynamics, the smoothness of the value function. Hence, we will prove the value function is the solution to a generalized free boundary problem. Using these results, we show how solutions of two relevant linked optimal stopping problems are found by solving two free boundary problems. The resulting optimal trading strategy derived from various plausible choices for price dynamics and the trader's utility function will be described and contrasted with the trading behaviour that arises when following the support/resistance line method. An example where all quantities are computed in closed form is presented and a further counterexample (where key assumptions are not satisfied) is derived.  
We develop further examples via a numerical approach and examine comparative statics with respect to risk aversion, the decay rate $\l$, the interest rate, and the strength of partial reflection at the support/resistance line. 

We emphasise that, in contrast to standard regime-switching models, the regime transition in our model is path-dependent and not specified by an exogenous Markov chain. The path-dependent regime-changing can be viewed as a novel method of introducing a market signalling effect into the price process (see Lehalle and Neuman \cite{Leh} for a different approach). 
We stress that the optimal stopping problems presented here are not standard since the stock price process is {\em not} a diffusion and the regime process on its own is {\em not} Markovian, unlike standard regime-switching models. 
We appreciate our model cannot reflect all the complexities of the path-dependencies of TA but we aim to take steps towards a better understanding in an interesting and challenging setting.

The vast majority of studies of TA devote their efforts to finding empirical evidence for the profitability of technical trading rules by examining historical data. For example, Brock, Lakonishok, and LeBaron \cite{Bro} tested moving-average-type trading rules and the support/resistance line method on the Dow Jones Industrial Average on a time scale of 90 years. This study suggested that the technical trading strategies considered there were significantly profitable. Based on a similar approach but with the data taken from Asian markets, Bessembinder and Chan \cite{Bes} further confirmed the forecasting power of trading rules based on TA. Lo, Mamaysky, and Wang \cite{Lo} implemented an automatic trading algorithm based on more sophisticated pattern-based trading rules (such as triangle, rectangle, and head-and-shoulders) by using kernel regressions, and a significant profit was observed. Park and Irwin \cite{Par} provided a comprehensive review of the literature on the profitability of TA and concluded that more than half showed positive evidence, though many of them had imperfections in their test procedures (for example, some ignored transaction costs). Ebert and Hilpert \cite{Ebe} demonstrated that the market timing of technical trading rules induced skewed trading profits. Popular rules were studied by a combination of simple models, simulations and analysis of empirical data. They argued that investors’ preference for positive skewness partially explained the popularity of TA. 
Tremendous effort has also been devoted to building algorithms which implement technical analysis-based trading strategies fast and accurately. For instance, Sezer, Ozbayoglu, and Dogdu \cite{Sez} designed a trading system based on a neural network constructed by using technical trading rules (based on the simple moving average and the relative strength index), and they showed the optimised system did outperform a buy-and-hold strategy. 

In contrast, very little research has been done on the mathematical modelling side. Blanchet-Scalliet et al. \cite{Bla} derived the optimal expected portfolio wealth at some terminal time $T$ where the underlying price process was assumed to have a mis-specified drift from time 0 to an exponentially distributed random time $\tau$, and (using Monte Carlo methods) they numerically  compared it with the expected portfolio wealth resulted from a simple moving-average trading strategy. Lorig, Zhou, and Bin \cite{Lor} studied a logarithmic utility maximization problem when trading strategies are based on exponential moving averages of the price of an underlying risky asset. De Angelis and Peskir \cite{Ang} determined the optimal stopping time that minimised the expected absolute distance between the stock price and the unknown support/resistance line which was assumed to be a random variable independent of the price. In a preliminary version of this paper, (Jacka and Maeda \cite{Jac}), and under an unrealistic constraint with only linear utility, two linked optimal stopping problems were solved with a model for the stock price inspired by the support/resistance line method. 
Nevertheless, this literature either focused on particular dynamics (e.g. De Angelis and Peskir \cite{Ang}) or a specific utility function (e.g. Blanchet-Scalliet et al. \cite{Bla} and Lorig, Zhou, and Bin \cite{Lor}, Jacka and Maeda \cite{Jac}).

The rest of this paper proceeds as follows. In Section \ref{preliminaries}, we provide definitions for key ingredients of the model and establish important mathematical properties. In Section \ref{opp}, we give some general results regarding the optimal stopping problem. In Section \ref{seller's problem}, we describe and solve \textit{the seller's problem} and obtain the optimal selling boundaries. An example where all quantities are computed in closed form is presented and a further counterexample (where key assumptions are not satisfied) is derived.  
In Section \ref{buyer's problem}, we define and solve \textit{the buyer's problem}, which provides the optimal buying boundaries and extend the explicit example from the previous section. In Section \ref{risk aversion}, we analyse numerically the influence of risk aversion on the optimal trading strategies and study comparative statics with respect to the decay rate $\l$ and the strength of partial reflection at $R$. 


\section{A path-dependent regime-switching model}\label{preliminaries}

Initially, we suppose that there are two price levels $L$ and $H$ (with $0<L<H$) and two regimes: positive (denoted $+$) and negative (denoted $-$) which are tracked by a flag process $F$. We assume there is a price level $R$ located in $(L,H)$ which is a support line if the stock price is in the positive regime and becomes a resistance line if the stock transitions into the negative regime. 
Then we extend the model by adding two more features: (i) an additional regime, denoted by $0$, such that $F$  transitions to $0$ from the + regime after an exponential waiting time, and in the 0 regime, $F$ can only transition to the negative regime by hitting $L$ from above; (ii) the support/resistance line $R$ is a partially reflecting barrier for the process, with reflection parameters $p_+ \in [0.5,1)$, $p_- \in (0,0.5]$ and $p_0 = 0.5$ (we use $q_f$ to denote $1-p_f$). 

More formally, letting $W$ be a Brownian motion then, given the flag process $F$ we take $S$ to solve
\begin{equation}\label{defS}
dS_t = \mu_{F_{t}}(S_t) dt + \sigma_{F_{t}}(S_t) dW_t+ (p_{F_{t}}-q_{F_{t}}) dl^R_t \text{,}
\end{equation}
where $l^R_t$ is the symmetric local time process of $S$ at $R$.

Then we take $F$ to be piecewise constant with jumps given by 
\begin{equation}\label{defF}
F_t = 
\begin{cases}
+ \quad \text{if $F_{t-}=-$, and $S_t=H$}  \\
0 \quad \text{if $F_{t-}=+$, and $t-T_n = J_n$ for some $n>0$} \\
- \quad \text{if $F_{t-}=+$, and $S_t=L$}\\
- \quad \text{if $F_{t-}=0$, and $S_t=L$}. 
\end{cases}
\end{equation}
where $T_n$ is the $n$th transition time of $F$ from $-$ to $+$ (with $T_1:=0$ if $F_0=+$)  and $(J_n)_{n\ge 1}$ is a sequence of i.i.d. Exponential($\l$) random variables independent of $W$.
Thus the regime transitions happen when: (i) $S$ hits $L$  from above and $F$ is in either the positive or zero regime, (ii) when the exponential clock goes off when in the positive regime, and (iii) when $S$ hits $H$ from below while $F$ is in the negative regime. It follows that the regime switching times are a sequence of stopping times which depend on the path of $S$. The existence of a process $(S,F)$ with these properties is guaranteed in Theorem \ref{fund} below. 
\begin{remark}
The symmetric local time is defined in Definition 5.47 of Jacod \cite{Jacod} p184 and the corresponding version of the Ito-Tanaka-Meyer formula is given in Theorem 5.5.2 on p186. The reason it is termed \lq symmetric local time' is that the symmetric versions of the $sign$ function:
$$
sign(x):=\begin{cases}
-1&x<0\\
0&x=0\\
1&x>0,
\end{cases}
$$
and of the derivative of a convex function $f$:
$$
f'(x)=\frac{1}{2}\bigl(f'_-(x)+f'_+(x)\bigr)
$$
(where $f'_-$ and $f'_+$ are the left- and right hand derivatives of $f$ respectively), are used in the 
definition of local time and in the Ito-Tanaka-Meyer formula.
\end{remark}
Note the notation $S^{x,f}$ will be used if we wish to emphasize the initial position. Where there is no fear of confusion, we will use $J$ (instead of $J_n$) to represent a generic holding time of $S$ in the positive regime before transitioning to the 0 regime. We assume $\l\ge 0$, and note that when $\l = 0$, $F$ never transitions from $+$ to $0$ and hence we restrict the statespace to $\R_+\times\{+,-\}$ and may view $F$ as only taking values in $\{+,-\}$.
 
With this set-up, the support/resistance effect of $R$ is directly introduced into the model. The impact of the additional $0$ regime is to prevent the process from remaining in the $+$ regime for a long time. The resulting statespace is $E =\bigcup_{f \in \{+,-,0\}} E^f \times \{f\}$ where $E^+ = E^0 =  (L,\infty)$ and $E^- = [0,H)$. Denoting the Euclidean metric on $\R$ by $\rho$, we define the metric $d$ on $E$ by
$$
d((x,f),(y,g))=\begin{cases}
\rho(x,y):&f=g\\
1+\rho(x,y):&f\neq g,
\end{cases}
$$
and denote the corresponding Borel $\sigma$-algebra by $\B$.

It is useful to separately define the three \lq component' diffusion processes $S^f$  which are solutions of the following SDEs,
\begin{equation}\label{stock}
dS^f_t = \mu_{f} (S^f_t)dt + \sigma_{f}(S^f_t)dW_t +(p_f-q_f) dl^R_t , \\
\end{equation}
with laws $\P^f_x$. Note that there is no reflection in the $0$ regime, since $p_0=q_0=0.5$. 

\begin{remark}Define $\q_f(x)=x1_{(x<R)}+ 1_{(x \geq R)}(R+\frac{q_f}{p_f}(x-R))$, so that $\q_f:\R_+\rightarrow\R_+$ is a continuous and strictly increasing bijection. We observe that $\q^f$ has symmetric derivative
$$
{\q_f}'_+=\begin{cases}
1&:x<R\\
\frac{1}{2p_f}&:x=R\\
\frac{q_f}{p_f}&:x> R,
\end{cases}
$$
and has a second derivative, in the sense of measures, given by $\rho(dx)=\frac{q_f-p_f}{p_f}1_{\{R\}}(x)$.
It follows that, defining the process $Y^f:=\q_f(S^f )$ and applying the Ito-Tanaka-Meyer  formula (Theorem 5.5.2 of \cite{Jacod}):
$$
dY^f_t=\q_f'(S^f_t)dS^f_t+\frac{1}{2}\int_{\R}l^a_t(S^f)\rho(da)=\q_f'(S^f_t)\sigma_f (S^f_t)dB_t+\q_f'(S^f_t)\mu_f(S^f_t)dt,
$$
or, writing $S^f_t$ as $\q_f^{-1}(Y^f_t)$,
\begin{equation}\label{sdef}
dY^f_t=\q_f'\circ \q_f^{-1}(Y^f_t)\sigma_f\circ \q_f^{-1}(Y^f_t)dB_t+\q_f'\circ\q_f^{-1}(Y^f_t)\mu_f\circ\q_f^{-1}(Y^f_t)dt.
\end{equation}
\end{remark}
To ensure the (weak) existence and uniqueness of $(S,F)$, we make the following assumptions on the dynamics: 
\begin{assumption}\label{dyn1}
$\sig_f:\R_+\rightarrow \R_+$ is a non-negative Borel-measurable function and $\mu_f:\R_+\rightarrow \R$ is Borel-measurable for each $f\in\{+,-,0\}$.
 
	Furthermore, defining 
	\begin{align*}
	&M := \{(x,f)\in E: \sig_{f}(x) = 0 \}\\ 
	\text{and}& \\
&N :=\{(x,f) \in E: \int_{N_x} \sig^{-2}_{f}(y) dy = \infty, \text{ for any open set $N_x$ in $\R_+$ containing $x$ } \},
\end{align*}
we have
	\begin{align}
	& M = N = \{(0,-)\},\label{int2} \\
	& \{(x,f) \in E: \int_{N_x} \biggl|\frac{\m_{f}(y)}{\sigma^2_{f}(y)}\biggr|dy 
	< \infty, \text{ for any open sets $N_x$ containing $x$ } \} = E\setminus N\label{int}.
	\end{align}
\end{assumption}
\begin{remark}Note that, since $|\frac{\q_f(x)}{x}|$ is bounded and bounded away from 0, Assumption \ref{dyn1} also applies to $\sig_f\circ\q_f^{-1}$ and $\mu_f\circ\q_f^{-1}$, so, applying Theorem 4.53 (2) 
of Engelbert and Schmidt \cite{Eng},  under Assumption \ref{dyn1}, there exist solutions of (\ref{sdef})  which are unique in law. Since $\q_f$ is a bijection we can conclude that the same applies to solutions of (\ref{stock}).
\end{remark}

To construct $(S,F)$ we need versions of $S^f$ , killed on hitting $L$ and $H$. 
We denote the infinitesimal generators of the killed processes by $\L^f$.
We denote the scale functions and speed measures in the three regimes by $s_f$ and $m_f$ respectively.
\begin{lemma}\label{scale}
Up to shift and scale changes, the scale function $s_f$ is given  by its symmetric derivative
\begin{equation}\label{scale1}
s_f'(x)=\begin{cases}
\exp\biggl(-\int_R^x\frac{2\mu_f(t)}{\sigma^2_f(t)}dt\biggr):&x<R\\
\frac{1}{2p_f}:&x=R\\
\frac{q_f}{p_f}\exp\biggl(-\int_R^x\frac{2\mu_f(t)}{\sigma^2_f(t)}dt\biggr):&x> R,
\end{cases}
\end{equation}
and the speed measure is given by
$$
m'_f(dx)=\frac{2}{s'_f(x)\sigma^2_f(x)}dx.
$$
\end{lemma}
\begin{proof}
This follows in the same way as (\ref{sdef}). We can see that $s_f$ is the difference of convex functions by decomposing $s_f'$ into its positive and negative parts and applying (\ref{int}). Then, applying the Ito-Tanaka-Meyer formula we see that
$$
ds_f(S^f_t)=\sigma_f(S^f_t)s_f'(S^f_t)dW_t,
$$
so that
$s_f(S^f)$ is a local martingale, as required. The well-posedness of the definition of $m_f$ follows from (\ref{int2}); its form follows either from a time-change argument or from Proposition 3.12 in Chapter VII of \cite{Rev}.
\end{proof}
\begin{remark}
We may also prove Lemma \ref{scale} by applying Exercise 3.20 in Chapter VII of \cite{Rev} to $\q_f(S^f)$.
\end{remark}
We assume: 
\begin{assumption}\label{dyn2}
	Khasminskii's condition holds in the positive and zero regimes:
	\begin{equation}\label{test}
	\int_{1}^{\infty}s_0(dx)\int_{1}^{x}m_0(dy)=
	\int_{1}^{\infty}s_+(dx)\int_{1}^{x}m_+(dy)=\infty.
	\end{equation}
\end{assumption}
\noindent This implies the processes $S^+$ and $S^0$ do not explode in finite time (see Rogers and Williams \cite{Rogers2} p.297), and hence $S$ will inherit this property. 

We will assume  Assumptions \ref{dyn1} and \ref{dyn2} are in force  in the rest of this paper.
From Assumption \ref{dyn1} the Ito diffusions $S^f$ are \textit{regular} except at 0 (i.e. $\P_x(S^f \ \text{hits}\ y)>0$,  for all $x>0$ and $y> 0$, for each $f \in \I:=\{+,0,-\}$).



\begin{thm}\label{fund}
Under Assumptions \ref{dyn1} and \ref{dyn2}: if $\l>0$, a (time-homogeneous) Markov process $(S,F)$ satisfying (\ref{defS}) and \ref{defF}) exists and is unique in law. For $g:E\rightarrow \R$, defining $g_f$ by $g_f:x\mapsto g(x,f)$, $(S,F)$ has infinitesimal generator $\L$ given by 
$$
\L g(x,f)=\L^f g_f(x)+\l (g(x,0)-g(x,+))1_{(f=+)},
$$
with $\D(\L)=\{g:\; g_f\in\D(\L^f)\text{ for each }f\in \I\}$.

The process $(S,F)$ is Feller, and thus has the strong Markov property, and is regular at all points in $E$ except $(0,-)$.
It follows from (\ref{defS}) that $S$ is a continuous semimartingale. Moreover $(S,F)$ is c\`adl\`ag.

If $\l=0$ then the same statements hold when we restrict $(S,F)$ to $\R_+\times\{+,-\}$.
\end{thm}

The central idea of the proof of Theorem \ref{fund} is that we can glue the laws of $S^+$, $S^0$, and $S^-$ together at the countable sequence of  stopping times corresponding to regime transitions. We present the  proof of Theorem \ref{fund} in Appendix \ref{fundproof}. 

From now on, we will work with a process $(S,F)$ which is defined on a  filtered probability space $(\Omega,\F,\bF=\{\F_t\}_{t \in \R_+\cup\{\infty\}}, \P_{x,f})$, satisfying the usual conditions, which supports a Brownian motion $W$. We stress that, by Assumption \ref{dyn1}, either $(0,-)$ is inaccessible or is absorbing for $(S,F)$. We will consider both cases and distinguish the results where they are different. In the case where $(0,-)$ is inaccessible (e.g. $S^-$ is a geometric Brownian motion), the statespace $E$ may exclude $(0,-)$, but we trust that there is no prospect of confusion in still using $E$ to denote it.  Note that, for $A \in \F$, we say $A$ a.s. (or $\P$-a.s.), if $\P_{x,f}(A) =1 $ for each $(x,f) \in E$. 
\begin{defn}
For any $A \in \B$, we define the hitting time of $A$ by $\t_A := \inf\{t \ge 0:\; (S_t,F_t) \in A\}$. If $A \in \B(\R_+)$, we denote by $\tau^f_A$  the first time $S$ enters $A$ while in the $f$ regime, so that $\tau^f_A=\tau_{A\times \{f\}}$. If $A=\{a\}$ for some $a \in \R_+$, we simply use $\tau^f_a$ to denote $\tau^f_A$, $f \in I$. Finally, we trust that there will be no confusion if, for $A \in \B(\R_+)$, we set $\tau_A:=\tau_{A\times I}=\tau^+_A\land \tau_A^0\land \tau_A^-$.
\end{defn}
\begin{remark}
 According to Kallenberg \cite{Kal}, since $(S,F)$ is a right-continuous adapted process, it is progressively measurable (i.e., $(S,F)$ restricted to $\Omega \times [0,t]$ is $\F_t \otimes \B[0,t]$-measurable for every $t \ge 0$) and Theorem 7.7 in \cite{Kal} then ensures $\t_A$ is a Markov time (a stopping time if $\t_A$ is finite a.s.) for any $A \in \B$.
\end{remark}
We note that, under $\P_{x,f}$, $S$ has the same law as $S^f$ until the first time that the regime changes. Theorem \ref{fund} identifies $\L$, the infinitesimal generator of $(S,F)$.
More generally, let  $\mathbb{L}$ denote the extended or {\em martingale generator} of $(S,F)$, i.e. for a measurable function $h$, if there is a measurable function $g$ such that, $\int_{0}^{t} |g(S_s,F_s)|ds<\infty$ a.s. and for each $(x,f) \in E$, 
\begin{equation}\label{mgen}
M_t:=h(S_t,F_t)-h(x,f)-\int_{0}^{t} g(S_s,F_s)ds
\end{equation}
is a local martingale under each $\P_{x,f}$, then we say $\mathbb{L} h =g$ and $h \in \mathcal{D}(\mathbb{L})$. Similarly, for $S^f$, we denote its martingale generator by $\mathbb{L}^f$. 

\begin{remark}
Many authors (see e.g. \cite{Rev}) restrict the domain of the martingale generator to functions for which (\ref{mgen}) is a true martingale.

Note that $\D(\L^f) \subset \D(\mathbb{L}^f)$ and  $\D(\L) \subset \D(\mathbb{L})$, by Proposition 1.7 in Chapter VII of \cite{Rev}. Furthermore, we can show the following equivalence: 
\begin{lemma}
	If  $h: E \rightarrow \R$ is in $\mathcal{D}(\mL)$ iff $h_f \in \mathcal{D}(\mL^f)$ for each $f\in\I$ and $\mL h(x,f)=\mL^fh_f(x)+\l(h_0(x)-h_+(x))\1_{(f=+)}$.
\end{lemma}
\begin{proof}
Let $h \in \mathcal{D}(\mathbb{L})$ and suppose that $S_0,F_0)=(x,f)$.Then there is a Borel-measurable function, $g$, such that 
\begin{equation}
	M_{t }:=h(S_{t},F_{t })-h(x,f)-\int_{0}^{t} g(S_s,F_s)ds
\end{equation}
is a local martingale. 

Now let $\T$ be the first time that the regime changes.
Then, in the case $f\neq +$, 
since $S^f$ is stopped at the boundary where the regime switches, we see that
\begin{equation}
M_{t \land \t}=N_{t}:=h(S^f_{t\land \t},f)-h(x)-\int_{0}^{t\land \t} g(S^f_s,f)ds,
\end{equation}
which shows $N$ is also a local martingale. Finally, $\int_{0}^{t\land \t} |g(S^f_s,f)|ds=\int_{0}^{t\land \t} |g(S_s,F_s)|ds<\infty$ a.s. and we can conclude that $\mL^fh_f=g_f$.
In the case that $f=+$ a similar argument applies, allowing for the transition to the zero regime at rate $\l$.
The reverse implication follows easily in a similar fashion. Given $g_f$ such that
$$
h_f(S^f_t)-h_f(x)-\int_0^t g_f(S^f_s)ds\text{ is a local martingale for each }f,
$$
we see that
$$
h(S_t,F_t)-h(x,f)-\int_0^t\biggl(g_{F_s}(S_s)+\l\bigl((g_0(S_s)-g_+(S_s)\1_{(F_s=+)}\bigr)\biggr)ds
$$
is a local martingale.
\end{proof}
\end{remark}

\begin{remark}
If we define the differential operator $L^f$ by $L^fg:x\mapsto\half\sigma_f^2(x)g''(x)+\mu_f(x)g'(x)$ then it is easy to see from the Ito-Tanaka-Meyer formula that if $g\in C^2[a,b]\cap C^0(\R_+)$ with $R\not\in[a,b]$ then
$ \mL^f g=L^fg$. More generally, Proposition 3.12 of Chapter VII of \cite{Rev} tells us that
\begin{equation}\label{sgen}
\mL^fg=\frac{d}{dm_f}\frac{d}{ds_f} g\text{ for }g\in\D(\mL^f),
\end{equation}
in the sense that 
$\frac{dg}{ds_f}$ exists and 
$$\frac{dg}{ds_f}(x_2)-\frac{dg}{ds_f}(x_1)=\int_{x_1}^{x_2}\mL^f(y)m_f(dy).
$$
Conversely, the Ito-Tanaka-Meyer formula tells us that if $\kappa$ is locally $m_f$-integrable on $(0,\infty)$ (i.e $\int_a^b|\kappa(t)| dm_f(t)<\infty \text{ for every }0<a<b<\infty$ then 
\begin{equation}\label{geneqn}
g:x\mapsto \int_a^x \int_a^y \kappa(t)dm_f(t))ds_f(y)\in\D(\mL^f)\text{ with }\mL^fg=\kappa.
\end{equation}
\end{remark}
Fix a positive constant $r$, which shall be understood as the interest rate in later sections. Then, we define \it fundamental solutions, \rm which we denote by $\phi^f_\gamma$ and $\psi^f_\gamma$ for (\ref{ode1}),  (\ref{ode2}) and (\ref{ode3}) as follows:
\begin{equation}\label{fundamental solution}
\psi^f_\gamma(x)=
\begin{cases}
\E^{x}[e^{-\gamma\tau^f_{c^f}}] & \textrm{if $x \le c^f$}\\
\frac{1}{\E^{c^f}[e^{-\gamma\tau^f_{x}}]} & \textrm{if $x > c^f$,}\\
\end{cases}
\quad
\phi^f_\gamma(x)=\begin{cases}
\frac{1}{\E^{c^f}[e^{-\gamma\tau_{x}}]} & \textrm{if $x \le c^f$}\\
\E^{x}[e^{-\gamma\tau_{c^f}}] & \textrm{if $x > c^f$,}\\
\end{cases}
\end{equation}
where $c^+=H$ and $c^-=c^0=L$. Note that $\psi^f_\gamma$ is strictly increasing and $\phi^f_\gamma$ is strictly decreasing. We drop the subscript $\gamma$ in this notation when $\gamma=r$.

Note that if we write $\mL=\frac{s_f'}{m_f'}\frac{d^2}{{ds_f}^2}$, then a unique solution to the Dirichlet problem 
\begin{equation}\label{dirich}
\mL w-\gamma w=\Phi\text{ with boundary conditions }w(k)=K,\; w(m)=M,
\end{equation}
exists by Theorem 8.3 of \cite{Gil}, since $\frac{m_f'}{s_f'}\mL$ is in divergence form (in natural scale) and $\frac{\gamma m_f'}{s_f'}\ge 0$.

Standard arguments then establish that fundamental solutions are solutions to the following generalised ODEs:
\begin{align} 
{\mL}^+ v-\gamma v=0 \label{ode1}, \\
{\mL}^- v-\gamma v=0  \label{ode2},\\
{\mL}^0 v-\gamma v=0  \label{ode3},
\end{align}
with suitable discontinuities in the first derivative (with respect to Lebesgue measure) at $R$ corresponding to the jump in $s_f'$.

\begin{remark}\label{strict}It is easy to show, by pathwise comparison with a Brownian motion, the stronger statements that $\frac{d\psi^f}{ds_f}>0$ and $\frac{d\phi^f}{ds_f}<0$,
which implies that
 $(\psi^f_\gamma)'>0$ and $(\phi^f_\gamma)'<0$ except possibly at $R$.
\end{remark}

\section{The optimal stopping problems}\label{opp}
We  study two problems in this paper. The first is called the {\em seller's problem}. In this problem, a trader initially holds the stock  and seeks a  selling time which gives the maximum gains (utility in this paper). The second one we term the {\em buyer's problem} : here a trader wants to maximise expected utility (gain) by first purchasing a stock and then selling it later. Both problems are formulated as optimal stopping problems, and we will present some general results here.  

We assume we have a {\em gains function} $h: E \rightarrow \R_+$ and an interest rate $r > 0$. We introduce the following assumptions. 
\begin{assumption}\label{SA1}
$h$ is independent of $F$ and  is $C^2$.
\end{assumption}

As we will see, the following assumption guarantees the finiteness of the value function for our optimal stopping problems.
\begin{assumption}\label{SA2}
	$\E^{(x,f)}\Big [\sup_{t \ge 0 }|e^{-rt}h(S_t)| \Big] < \infty$.
\end{assumption}

The \it optimal stopping problem \rm is defined by 
\begin{equation}\label{prob}\tag{P}
\V(x,f):=\sup_{\tau}\E^{(x,f)}[e^{-r\tau}h(S_\tau)],
\end{equation}
where the supremum is taken over all stopping times, and we call $\V(x,f)$ the \textit{value function}. We also look for the optimal stopping time $\t^*$ making
\begin{equation}\label{stoptime}
\V(x,f)=\E^{(x,f)}[e^{-r{\tau^*}}h(S_{\tau^*})].
\end{equation}
\begin{remark}
The supremum in (\ref{prob}) is taken over all (finite) stopping times, but the value function $\V$ remains the same even if the supremum is taken over Markov times by Theorem 1 in Chapter 3  of \cite{Shir}, provided that we define

\begin{equation}\label{limsup}
e^{-r\tau}h(S_\tau)1_{(\tau=\infty)}:=\limsup_t e^{-rt}h(S_t)1_{(\tau=\infty)}
\end{equation} 
i.e. we set $e^{-r\tau}h(S_\tau)=\limsup_t e^{-rt}h(S_t)$ on the event $(\tau=\infty)$.
\end{remark}

The following lemmas are required for the proof of Theorem \ref{optimal stopping times}.

\begin{lemma}\label{UI}
Under Assumption \ref{SA2}, $\V(x,f)<\infty$ for any $(x,f) \in E$.
\end{lemma}
\begin{proof}
By definition,
\begin{equation}
	\V(x,f)=\sup_{\tau}\E^{(x,f)}[e^{-r\tau}h(S_\tau)] \le \E^{(x,f)}\Big [\sup_{\t}e^{-r\t}h(S_{\t}) \Big]= \E^{(x,f)}\Big [\sup_{t}e^{-rt}h(S_{t}) \Big]
\end{equation}
 Hence, by Assumption \ref{SA2}, $\V(x,f)< \infty$.
\end{proof}
\begin{lemma}\label{lsc}
Under Assumption \ref{SA1} and \ref{SA2}, the value function $\V(x,f)$ is continuous \\
$\big(i.e. \ \liminf_{y \rightarrow x} \V(y,f) \ge \V(x,f) \big)$.
\end{lemma}
\begin{proof}
By Assumption \ref{SA2}, we can apply Theorem 1 in Chapter 3  of \cite{Shir} to see $e^{-rt}\V$ is the smallest excessive majorant of the gains function $e^{-rt}h$. Then by Theorem 5 in Chapter 3  of \cite{Shir}, since  $(S,F)$ is a Feller process and $h$ is bounded below by $0$, $e^{-rt}\V$ is continuous and thus so is $\V$.
\end{proof}

Define the \textit{stopping set} $D$ and \textit{continuation set} $C$ by
\begin{equation}\label{D}
D=\{(x,f) \in E| \V(x,f) = h(x) \},
\end{equation}
\begin{equation}\label{C}
C=\{(x,f) \in E| \V(x,f) > h(x) \}.
\end{equation}
As $\V$ is continuous, $D$ is closed and $C$ is open. The following theorem follows immediately from Shiryaev \cite{Shir} Chapter 3, Theorem 3. 
\begin{thm}\label{optimal stopping times}
For any gains function $h$ satisfying Assumptions \ref{SA1} and \ref{SA2},  the Markov time $\t_D$ is optimal in the sense that equation (\ref{stoptime}) holds, and if $\t_D < \infty$ a.s. for every $(x,f) \in E$ then it is an optimal stopping time. 
\end{thm}

By a well-known result (e.g. see Jacka and Norgilas \cite{Nor} Theorem 2.10), $e^{-rt}\V(S_t,F_t)$ is the {\em Snell envelope} of  the gains process $e^{-rt}h(S_t)$ under Assumption \ref{SA2}, i.e. $e^{-rt}\V(S_t,F_t)=\esssup_{\t \ge t} \E[e^{-r\t}h(S_\t)|\F_t]$, a.s. Moreover, standard theory of optimal stopping (e.g. Theorem 2.2 of \cite{Pes}) tells us that $\bigl(e^{-rt}\V(S_{t},F_{t})\bigr)_{t\geq 0}$ is a supermartingale and the stopped process $\bigl(e^{-rt\land\t_D }\V(S_{t\land\t_D,F_{t\land\t_D})}\bigr)_{t\geq 0}$ is a martingale. 
Assumption \ref{SA2} ensures that $\bigl(e^{-rt}\V(S_{t},F_{t})\bigr)_{t\geq 0}$ is Class D.

\begin{lemma}\label{P123}
	Let $\func$ denote a function on $E$. Define $N_t=N^{x,f}_t := e^{-rt}\func(S_{t},F_{t})$ with $(S_0,F_0)=(x,f)$. For any $(x,f)\in E$, if $N^{x,f}$ satisfies properties P1,P2 and P3 defined as follows:
	\begin{enumerate}[(P1)]
		\item $N_t$ is a class D supermartingale,
		\item there is a Markov time $\tau $ such that $N_0=\mathbb{E}^{x,f}[e^{-r\tau}h(S_{\tau})]$,
		\item $N_t \ge e^{-rt}h(S_t)$ for all $t \ge 0$
	\end{enumerate}
	then, $\func(x,f)=\V(x,f)$. 
\end{lemma}
\begin{proof}
    By the Optional Sampling Theorem for  class D supermartingales (see Rogers and Williams \cite{Roger} pp.189), for any stopping time $\tau$,
	\begin{equation}\label{train}
	\func(x,f)=N_0\ge\E^{(x,f)}[e^{-r  \t}\func(S_{ \t}, F_{  \t})]\ge\E^{(x,f)}[e^{-r \t}h(S_{\t})],
	\end{equation}
	where the last inequality follows from P3. Since (\ref{train}) holds for any $\t$, we get $V(x,f) \ge \V(x,f)$. On the other hand, by P2, for some Markov time $\tau$,
	\begin{displaymath}
	\func(x,f)=N_0=\E^{(x,f)}[e^{-r\tau}h(S_\tau)],
	\end{displaymath}
	and hence $\func(x,f) \le \V(x,f)$.
\end{proof}
\begin{remark}\label{P2}Of course, if $N^{x,f}$ satisfies P2 then $\func(x,f)\leq\V(x,f)$.
\end{remark}
If a process $X$ starts at $x$ in the boundary of the continuation region, $C$ and enters int($D$) immediately with positive probability, then the smooth pasting principle is often valid at $x$ (see Section 9 in Peskir and Shiryaev \cite{Pes}). The smooth pasting principle is well established for Ito diffusion processes (see e.g., Jacka and Norgilas \cite{Nor}), but not in greater generality. Nevertheless, as we shall see, since the process $S$ is an Ito diffusion before the first regime transition, a suitable version of the smooth pasting principle indeed holds.

To solve the free boundary problem, we look for a measurable function $v:E \rightarrow \R$ and a set $\tD$ such that $v \in \D(\mathbb{L})$ and, denoting the boundary of an open set $U$ by $\partial U$, 
\begin{align}\mathbb{L} v-rv=0 \text{ in $\tC\setminus\{R\}\times\{+,-\}$, } \label{free1}\\	v |_{\tD}=h|_{\tD}, \label{free2}\\	\frac{\partial v}{\partial x} |_{\partial \tC}=\frac{\partial h}{\partial x} |_{\partial \tC}, \label{free3}
\end{align}
where $\tC:=\tD^c$. 

We shall see that the value function $\V$ and stopping set $D$ is a solution to a {\em free boundary problem}. 
We will show that, conversely, for our two problems the value function is the unique solution to the free boundary problem, under some extra conditions, and is the maximal solution regardless. This is done in Section \ref{sol1} for the seller's problem and in Section \ref{buyer's problem} for the buyer's problem.

\section{The seller's problem} \label{seller's problem}

\subsection{The Seller's Problem}\label{subsec: seller}
We restrict our attention to the class of gains function of the form $u:\R_+ \rightarrow \R_+$ such that 
\begin{assumption}[]~\label{assumption: utility}
	\begin{enumerate}
		\item $u \in C(\R_+) \cap C^2(0,\infty)$.
		\item $u$ has strictly positive derivative and is positive on $(0,\infty)$. 
		\item\label{sup} For each $(x,f)\in E$, $\E^{(x,f)}\Big [\sup_{t \ge 0 }|e^{-rt}u(S_t)| \Big] < \infty$.
		\item\label{limit} For each $(x,f)\in E$, $\E^{(x,f)}\Big [\limsup_{t \ge 0 }e^{-rt}u(S_t) \Big] =0$
	\end{enumerate}
\end{assumption}

The optimal stopping problem is, 
\begin{equation}\label{SP}
\V(x,f):=\sup_{\text{Markov times }\tau}\E^{(x,f)}[e^{-r\tau}u(S_\tau)]. 
\end{equation}

We recall that the stopping set is denoted by $D$ and the continuation set by $C$ . We conclude that $\t_D$ is optimal by Theorem \ref{optimal stopping times}, and $e^{-rt}\V(S_t,F_t)$ is the Snell envelope of  $e^{-rt}u(S_t)$. 
\begin{remark}\label{Limsup}
Since $u\ge 0$, Assumption \ref{assumption: utility} \ref{limit} implies that 
\begin{equation}\label{limsup1}
\limsup e^{-rt}u(S_t)=\lim e^{-rt}u(S_t)=0\text{ a.s.}
\end{equation}
\noindent Moreover, by Lemma 12 in Chapter 3 of \cite{Shir}, this implies that 
\begin{equation}\label{limsup2}
\limsup e^{-rt}\V(S_t)=\lim e^{-rt}\V(S_t)=0\text{ a.s.}
\end{equation}

\end{remark}

\begin{assumption}\label{c}
	There is a constant $A > H$ such that
	\begin{IEEEeqnarray}{rl}
		\mL^- u-ru < 0 &\quad \textrm{in} \quad (0, H), \label{u in -}\\
		\mL^+ u-ru > 0 &\quad \textrm{in} \quad (L, A), \label{u in L-A}\\
		\mL^+ u-ru < 0 &\quad \textrm{in} \quad (A, \infty),\label{u in A - infty}\\
		\mL^0 u-ru < 0 &\quad \textrm{in} \quad (L, \infty).\label{u in 0}
	\end{IEEEeqnarray}
\end{assumption}

\begin{remark}\label{remark2}
	For some common choices of dynamics and utility functions (e.g. a geometric Brownian motion with a power utility function), $\mL^+ u-ru\equiv \half \sigma_+^2x^2 u''+\mu_+xu'-ru$ can only have one sign. In this case, the optimal stopping time (in the positive regime) can be proven to be either $0$ or $\infty$, which is neither very interesting nor realistic. We observe that our assumptions {\em are} satisfied for a wide class of realistic dynamics and utility functions. 
\end{remark}
\begin{remark}\label{remark3}
	For the geometric Brownian motion, there are a wide range of utility functions which satisfy Assumption \ref{c}. For example, consider an exponential utility function $u(x)=\frac{1-e^{-ax}}{a}$ for $a > 0$. Then,
	\begin{equation}
		\mL^f u - ru = e^{-ax}(-\half \sig_f x^2a+\m_fx+\frac{r}{a}) -\frac{r}{a}.
	\end{equation}
	For $a\ge 1$, we can choose $\sig_f$ and $\m_f$ such that Assumption \ref{c} holds. 
\end{remark}

\begin{remark}\label{remark4}
	Note that neither does $u$ being a utility function imply Assumption \ref{c} nor does the reverse implication hold. 
\end{remark}

\begin{assumption}\label{suffa}
Suppose $u$ satisfies 
\begin{equation}\label{suff}
\limsup_{x\rightarrow\infty}\max\biggl[\biggl(\frac{\sigma_0(x)u'(x)}{u(x)}\biggr)^2,\biggl(\frac{\sigma_+(x)u'(x)}{u(x)}\biggr)^2\biggr]<\infty,
\end{equation}
and
\begin{equation}\label{suff2}
\exists\epsilon>0,\Delta\text{ such that }\max\bigl[\mL^+u(x)-(r-\epsilon)u(x),\mL^0 u(x)-(r-\epsilon)u(x)\bigr]\leq 0\text{ for }x\geq \Delta.
\end{equation}
\end{assumption}

\begin{thm}\label{condition}
Suppose that, in addition to Assumption \ref{c}, $u$ satisfies Assumption \ref{suffa}.  
Then $u$ satisfies conditions \ref{sup} and \ref{limit} of  Assumption \ref{assumption: utility}.
\end{thm}
\noindent See Appendix \ref{addl} for the proof.

\subsubsection{Boundaries of stopping sets}
We will show 
 \begin{thm}[]~\label{theorem: D shape}
 	\begin{enumerate}
 		\item Defining	$B:= \inf\{x\in D^+\}$, then $B\geq A$.
 		\item   $D^-=[0,m]$ (if $0$ is absorbing) or $(0, m]$ (if $0$ is inaccessible) for some $m \in [0,H)$.
 		\item   If $m \ge L$, then $D^0 = (L,\infty)$. If $m < L$, then there exists constant $c \in (L,\infty)$ such that $D^0 = [c,\infty)$.
 		 		\item\label{last} If (i) $m\ge L$ or if (ii) $m< L$ but $B\geq c$ or if (iii) $\l = 0$, then $D^+=[B,\infty)$.
 		\item If  $m<L$ and $B<c$ and $\l>0$ then either $D^+\supseteq [c,\infty)$ or, defining $\r=\sup C^+$ and $\lef=\sup (D^+\cap(L,c))$,
 		$$c<\r \text{ and }(\lef,\r)\subset C^+.$$
If Assumption \ref{suffa} holds then $\r<\infty$.
 	\end{enumerate}
 \end{thm}
 
\begin{proof}
\begin{enumerate}
\item Suppose $\exists y \in (L,A)$ such that $y \in D^+$. Then there is $\epsilon >0$ with $L+\epsilon<y < A-\epsilon $. Define $\tau=\tau^+_{L+\epsilon} \land \tau^+_{A-\epsilon}\land J$.  By the Ito-Tanaka-Meyer  formula, 
$$
e^{-r({t\wedge \tau})}u(S_{t\wedge \tau})=u(y)+M_{t\wedge \tau}+\int_0^{t\wedge \tau} e^{-rs}(\mL^+ u(S_s)-ru(S_s)) ds+(p_+-q_+)u'(R)\int_0^{t\wedge \tau} e^{-rs}dl_s,
$$
where $M_t=\int_0^t \sigma(S_t)u'(S_t)dW_t$. Assumption \ref{c} (more specifically (\ref{u in L-A})) now tells us that $e^{-r({t\wedge \tau})}u(S_{t\wedge \tau})$ is a positive submartingale.
Then, since $e^{-rt}u(S_t)$ is class D by property \ref{sup} of Assumption \ref{assumption: utility}, $M$ is uniformly integrable and so we may take expectations to conclude that
\begin{IEEEeqnarray}{rl}\label{dplus}
\E^{(y,+)}[e^{-r\tau}u(S_\tau)]&=u(y)+\mathbb{E}^{(y,+)}\Big[\int_0^\tau e^{-rt}(\mL^+ u(S_t)-ru(S_t)) dt+(p_+-q_+)u'(R)\int_0^\tau e^{-rt}dl_t\Big]\nonumber\\
& > u(y)=\V(y,+),
\end{IEEEeqnarray}
the strict inequality in (\ref{dplus}) following from property \ref{sup} of Assumption \ref{assumption: utility} and the fact that $\tau>0$ a.s.
But this contradicts the definition of $\V$. Thus, $D^+ \cap (L,A) = \emptyset$. 
\item To prove $D^-$ is an interval, suppose not: then there are  $y_1, y_2 \in D^-$ such that $H > y_2 > y_1 > 0 $ and $(y_1,y_2) \subset C^- $. Take any $y \in (y_1, y_2)$ and define $\tau=\tau^-_{y_1} \land \tau^-_{y_2}$. 
Since $\tau$ is an optimal Markov time, if $S$ starts at $(y,0)$,
$$
\V(y,-)=\E^{(y,-)}[e^{-r\tau}u(S_\tau)].
$$

The Ito-Tanaka-Meyer  formula gives
$$
e^{-r({t\wedge\tau})}u(S_{t\wedge\tau})=u(y)+\int_0^{t\wedge\tau}e^{-rs}(\mL^- u(S_s)-ru(S_s)) ds-(q_--p_-)u'(R)\int_0^{t\wedge\tau} e^{-rs}dl_s+M_{t\wedge\tau},
$$
where $M_t=\int_0^t \sigma(S_s)u'(S_s)dW_s$. Moreover,  since $\V$ is a class D supermartingale, $M$ is a uniformly integrable martingale (by the Doob-Meyer Supermartingale Decomposition Theorem) so, by the Optional Sampling Theorem
\begin{IEEEeqnarray}{rl}\label{dmin}
\V(y,-)&=\E^{(y,-)}[e^{-r\tau}u(S_\tau)]\nonumber\\
&=u(y)+\mathbb{E}^{(y,-)}\Big[\int_0^\tau e^{-rt}(\mL^- u(S_t)-ru(S_t)) dt-(q_--p_-)u'(R)\int_0^\tau e^{-rt}dl_t\Big]\nonumber\\
&\le u(y),
\end{IEEEeqnarray}
where $l$ denotes the local time at $R$ of $S$ and the last inequality in (\ref{dmin}) follows from property (\ref{u in -}) of Assumption \ref{c}.
But this contradicts the inequality $\V(y,-) > u(y)$, which holds because $y \in C^-$.
Therefore, $D^-$ is a closed interval.\\
Moreover, if $(0,-) \in E$, then, by property \ref{limit} of Assumption \ref{assumption: utility}, if $(S_0,F_0)=(0,-)$ we must have $e^{-r\tau}u(S_{\tau})=e^{-r\tau}u(0) \le u(0)$ a.s. for all Markov times $\tau$. Hence, 
$\E^{(0,-)}[\sup_{\tau}e^{-r\tau}u(S_{\tau})]\le u(0)$, which implies $0 \in D^-$.\\
So either $D^-=[0,m]$ for some $m \in [0,H)$ or $D^-=[0,H)$. To rule out the latter possibility, assume that it is true. Then $\forall \epsilon > 0$, $\V(H-\epsilon,-)=u(H-\epsilon)$, which implies $\lim_{\epsilon \rightarrow 0}\V(H-\epsilon,-)=u(H)$. However, because $e^{-rt}\V(S_t,F_T)$ is a class D positive supermartingale,
\begin{IEEEeqnarray*}{rl}
\V(H-\epsilon,-)&\ge \E^{(H-\epsilon,-)}[e^{-r({\tau_H\wedge\tau_{H/2})}}\V(S_{\tau_H\wedge\tau_{H/2}})]\\
&\ge\V(H,+)\E^{(H-\epsilon,-)}[e^{-r\tau_H}\1_{\tau_H < \tau_{H/2}}]+u(H/2)\E^{(H-\epsilon,-)}[e^{-r\tau_{H/2}}\1_{\tau_{H/2}< \tau_H}].
\end{IEEEeqnarray*}
Sending $\epsilon$ to 0, we can see that  $\lim_{\epsilon \rightarrow 0}\V(H-\epsilon,-) \ge \V(H,+) > u(H)$ as $\E^{(H-\epsilon,-)}[e^{-r\tau_H}\1_{\tau_H < \tau_{H/2}}]$ converges to $1$ and $\E^{(H-\epsilon,-)}[e^{-r\tau_{H/2}}\1_{\tau_{H/2}< \tau_H}]$ converges to 0 by continuity of $\phi_-$ and $\psi_-$. Therefore, by contradiction, $D^- \ne [0,H)$.\\
Next, suppose $(0,-) \notin E$. This implies $D^- = (0, m]$ or $D^- = [k,m]$ for some $k>0$. Assume $D^- = [k,m]$ and fix $x \in (0,k)$. Then $\tau^-_k$ is an optimal Markov time and $\V(x,-) >u (x)$ since $x \in C^-$, so
$$
\V(x,-)=\E^{(x,-)}[e^{-r{\tau^-_k}}u(S_{{\tau^-_k}})]> u(x).
$$
The Ito-Tanaka-Meyer formula,
gives, as before,
$$
e^{-r({t\wedge{\tau^-_k}})}u(S_{t\wedge{\tau^-_k}})=u(y)+\int_0^{t\wedge{\tau^-_k}}e^{-rs}(\mL^- u(S_s)-ru(S_s)) ds-(q_--p_-)u'(R)\int_0^{t\wedge{\tau^-_k}} e^{-rs}dl_s+M_{t\wedge{\tau^-_k}},
$$
where $M_t=\int_0^t \sigma(S_s)u'(S_s)dW_s$.  Property \ref{u in -} of Assumption \ref{c}, the assumption that $p_-\leq 0.5$, and property \ref{sup} of Assumption \ref{assumption: utility} imply that $e^{-r({t\wedge{\tau^-_k}})}u(S_{t\wedge{\tau^-_k}})$ is a class D supermartingale and thus
\begin{IEEEeqnarray*}{rl}
\V(x,-)&=\E^{(x,-)}[e^{-r{\tau^-_k}}u(S_{{\tau^-_k}})]\\
&=u(x)+\mathbb{E}^{(x,-)}\Big[\int_0^{{\tau^-_k}} e^{-rt}(\mL^- u(S_t)-ru(S_t)) dt-(q_--p_-)u'(R)\int_0^{\tau^-_k} e^{-rt}dl_t\Big]\\
&\le u(x),
\end{IEEEeqnarray*}
which leads to a contradiction. Thus, we must have $D^- = (0, m]$. 

\item We first show $D^0$ is an interval. Assume not, then there exist $L<a<b<\infty$ s.t $(a,b)\subseteq C^0$ with $a,b\in D^0$.
Take $y\in(a,b)$ and set $\tau=\tau^0_{\{a,b\}}$, then the Ito-Tanaka-Meyer formula tells us that
$$e^{-r\tau}u(S^0_\tau)=u(y) +M_\tau+\int_0^\tau e^{-rt}[\mL^0u(S^0_t)-ru(S^0_t)]dt,
$$
where $M_t=\int_0^t \sigma(S^0_t)u'(S^0_t)dW_t$.
Since $u$ is bounded on $[a,b]$, Assumption \ref{c} (\ref{u in 0}) tells us that $e^{-r({t\wedge\tau})}u(S^0_{t\wedge\tau})$ is a class D supermartingale, so that $M$ is a uniformly integrable martingale and we can take expectations to conclude that
$$
u(y)<\V(y,0)=\E^{y}[e^{-r\tau}u(S^0_\tau)]=u(y)+\E^{y}[\int_0^{\tau}e^{-rt}[\mL^0u(S^0_t)-ru(S^0_t)]dt]<u(y)
$$
(the last, strict,  inequality following from Assumption \ref{c} (\ref{u in 0}) and the fact that $\tau>0$ a.s.)
This is a contradiction, so we conclude that $D^0$ is an interval.

To show that $\sup D^0=\infty$, suppose not, then
$$
(a,\infty)\subseteq C^0,
$$
for some $\infty>a\geq L$.
Suppose first that $a=L$, then $D^0$ is empty and so if $(S,F)$ starts at $(y,0)$, with $y>B$, we wait first until $(S,F)$ hits $(L,-)$ and then until it eventually hits either $(m,-)$ or $(B,+)$. But, since $y>B$ and $u$ is increasing, this gives a payoff less than $u(y)$ which contradicts the assumption that $(y,0)\in C^0$. 

Now suppose that $a>L$,  and $(S,F)$ starts at $(y,0)$ with $y>a$. Then $\tau^0_a$ is an optimal Markov time, and, as above, $e^{-r({t\wedge\tau^0_a})}u(S^0_{t\wedge\tau^0_a})$ is a class D supermartingale
converging a.s. to 0 on $(\tau^0_a=\infty)$ so that
$$
\V(y,0)>u(y)>u(a)\geq \E^{(y,0)}[e^{-r({\tau^0_a})}u(S^0_{\tau^0_a})],
$$
 but this contradicts the optimality of $\tau^0_a$ and we conclude that, since $D^0$ is closed in $E^0$,  either $D^0=(L,\infty)$ or $D^0=[c,\infty)$ for some $c>L$. 
 
 To show that $D^0=(L,\infty)$ iff $m\ge L$, suppose first that $m<L$. Then $\V(L,-)>u(L)$ and, by continuity, $\lim_{x\downarrow L}\V(x,0)>u(L)$. Thus $\V(x,0)>u(x)$, and so $x\in C^0$, for $x$ sufficiently close to $L$. Conversely, if $m\ge L$ and $D^0=[c,\infty)$, take an $x\in(L,c)$. Then $\V(x,0)>u(x)$ and, starting at $(x,0)$, the Markov time $\tau^-_L\wedge \tau^0_c$ is optimal, so that $\V(x,0)=\E^{(x,0)}[e^{-r({\tau^-_L\wedge \tau^0_c})}\V(S_{\tau^-_L\wedge \tau^0_c})]$. But, by a now standard argument, $e^{-r({t\wedge\tau^-_L\wedge \tau^0_c})}\V(S_{t\wedge\tau^-_L\wedge \tau^0_c})$ is a class D supermartingale, so that
 $u(x)\ge \E^{(x,0)}[e^{-r({\tau^-_L\wedge \tau^0_c})}\V(S_{\tau^-_L\wedge \tau^0_c})]$, giving a contradiction.

\item	Next, let us prove $D^+$ is connected if either (i) $m\ge L$ or (ii) $c\leq B$. Suppose (to seek a contradiction) there exists a $y_1$ with $A\le B\le y_1\in D^+$, such that $(y_1, y_1+\epsilon) \subset C^+$ for some $\epsilon>0$.
Now define $y_2=\inf\{x>y_1:\; x\in D^+\}$, with $\inf\emptyset:=\infty$,
and then define $\tau=\tau_{y_1} \land \tau_{y_2}\land J$. Since either $c\le B$ or $m\ge L$ (in which case $D^0=(L,\infty)$), $(x,0)\in D^0$ for every $x\in(y_1,y_2)$ and it follows that, if $(S_0,F_0)=(y,+)$ for some $y\in(y_1,y_2)$, then $\tau$ is an optimal Markov time.
Thanks to property \ref{u in A - infty} of Assumption \ref{c} and, by now standard arguments,
	\begin{equation}\label{eq: D+connected}
	\begin{split}
	\V(y,f) & = \E^{(y,f)}[e^{-r\tau}u(S_\tau)] \\
	& = u(y)+\mathbb{E}^{(y,f)}\Big[\int_0^\tau e^{-rt}(\mL^+ u(S_t)-ru(S_t)) dt\Big]  \\
	& < u(y)
	\end{split}
	\end{equation}
	where the equality follows the optimality of $\t$ and from the fact that, by time $\t$, the process has not hit the partial reflection boundary $R$. 
	
	Finally, to show $D^+$ is connected if $\l = 0$, we notice (\ref{eq: D+connected}) still holds with $\t$ changed to $\tau=\tau_{y_1} \land \tau_{y_2}$. So we can argue analogously. This completes the proof. 
 	\item Suppose that $B<c<\r$. Take $y\in C^+$ with $y\geq c$ and let the component of $C^+$ containing $y$ be $(a,b)$ then, if $a\geq c$, we can deduce a contradiction as in the proof of  \ref{last}. Then this must be the last component of $C^+$ and so must be of the form $(a,\r)$ with $a<c$. 
 	
To show that $\r<\infty$ under assumption \ref{suffa}, assume otherwise. Then for any $z>c$, $z\in C^+\cap D^0$. It follows that for such a $z$, if $(S_0,F_0)=(z,+)$ then $\tau^+_d\land \tau_{E^0}\leq \tau_D$ for any $c\leq d\le z$.
Now $e^{-r({t\land \tau_D})}\V(S_{t\land \tau_D},F_{t\land \tau_D})$ is a uniformly integrable martingale , so for such a choice of $d$ and $z$:
\begin{IEEEeqnarray}{rl}\label{bigv}
\V(z,+)&=\E^{(z,+)}[e^{-r({\tau^+_d\land \tau_{E^0}})}\V(S_{\tau^+_d\land \tau_{E^0}},F_{\tau^+_d\land \tau_{E^0}})]\nonumber\\
&=\frac{\phi^+_{\l+r}(z)}{\phi^+_{\l+r}(d)}\V(d,+)+\E^{(z,+)}[e^{-r\tau_{E^0}}u(S^0_ {\tau_{E^0}})1_{( \tau_{E^0}<\tau^+_d)}]\label{big},
\end{IEEEeqnarray}
since, by equation (\ref{limsup2}) of Remark \ref{Limsup}, $e^{-r({\tau^+_d\land \tau_{E^0}})}\V(S_{\tau^+_d\land \tau_{E^0}},F_{\tau^+_d\land \tau_{E^0}})=0$ on the event  $({\tau^+_d\land \tau_{E^0}}=\infty)$
Taking $d\geq c$,
we see that 
\begin{IEEEeqnarray}{rl}
\E^{(z,+)}[e^{-r{\tau_{E^0}\land \tau^+_d}}u(S_ {\tau_{E^0}\land \tau^+_d})]
&=\E^{(z,+)}[e^{-r\tau_{E^0}}u(S_ {\tau_{E^0}})1_{( \tau_{E^0}<\tau^+_d)}]+\E^{(z,+)}[e^{-r\tau^+_d}u(d)1_{( \tau_{E^0}>\tau^+_d)}]\nonumber\\
&=\frac{\phi^+_{\l+r}(z)}{\phi^+_{\l+r}(d)}u(d)+\E^{(z,+)}[e^{-r\tau_{E^0}}u(S^0_ {\tau_{E^0}})1_{( \tau_{E^0}<\tau^+_d)}]
\end{IEEEeqnarray}
Conversely, applying the Ito-Tanaka-Meyer formula, and using the fact that $e^{-r({t\land \tau^+_c})}u(S^+_{t\land \tau^+_c})$ is a class D positive supermartingale,
\begin{IEEEeqnarray}{rl}\label{bigu2}
&\E^{(z,+)}[e^{-r{\tau_{E^0}\land \tau^+_d}}u(S_ {\tau_{E^0}\land \tau^+_d})]
=u(z)+\E^{(z,+)}[\int_0^{\tau_{E^0}\land \tau^+_d}e^{-rt}(\mL^+u-ru)(S_t)dt]\nonumber\\
&\Rightarrow\nonumber\\
&u(z)=\frac{\phi^+_{\l+r}(z)}{\phi^+_{\l+r}(d)}u(d)+\E^{(z,+)}[e^{-r\tau_{E^0}}u(S^0_ {\tau_{E^0}})1_{( \tau_{E^0}<\tau^+_d)}]\nonumber\\
&\phantom{444444444444444444}-\E^{(z,+)}[\int_0^{\tau_{E^0}\land \tau^+_d}e^{-rt}(\mL^+u-ru)(S_t)dt]
\end{IEEEeqnarray}
It follows that, subtracting (\ref{bigu2}) from (\ref{bigv}),
\begin{equation}\label{v-u2}
0<\V(z,+)-u(z)\le \frac{\phi^+_{\l+r}(z)}{\phi^+_{\l+r}(d)}(\V(d,+)-u(d))+\E^{(z,+)}[\int_0^{\tau_{E^0}\land \tau^+_d}e^{-rt}(\mL^+u-ru)(S_t)dt].
\end{equation}
Then, taking $d$ larger than $\Delta$ defined in Assumption \ref{c},
\begin{equation}\label{v-u}
0<\V(z,+)-u(z)\le \frac{\phi^+_{\l+r}(z)}{\phi^+_{\l+r}(d)}(\V(d,+)-u(d))-\epsilon u(d)\E^{(z,+)}[\int_0^{\tau_{E^0}\land \tau^+_d}e^{-rt}dt].
\end{equation}
Now $\lim_{z\rightarrow \infty}{\phi^+_{\l+r}(z)}=0$ while 
$$\lim_{z\rightarrow \infty}\E^{(z,+)}[\int_0^{\tau_{E^0}\land \tau^+_d}e^{-rt}dt]=\lim_{z\rightarrow \infty}\E^{(z,+)}[\int_0^{\tau^+_d}e^{-(\l+r)t}dt]=\frac{1}{\l+r},
$$
so, taking sufficiently large $z$, the right hand side of (\ref{v-u}) is negative, which is a contradiction.
 	\end{enumerate}
\end{proof}

\begin{cor}
If $\V(\cdot,+)\geq \V(\cdot,0)$ then $D^+=[B,\infty)$
\end{cor}
\begin{proof}
Since $u(B)=\V(B,+)\geq \V(B,0)$ we see that $c\leq B$ and the result follows.
 	\end{proof}

We shall see that, at least if the free boundary point $m$ is not equal to $R$, then the smooth pasting conditions hold because, heuristically, the smoothness at the free boundary is a local property. Recall that we denote the boundary of an open set $U$ in $\R$ by $\partial U$.
\begin{thm}[]~\label{theorem: smoothpasting}[Smooth Pasting]
	\begin{enumerate}
	\item If $y\in \partial C^+$ then $\V_x(y, +) = u'(y)$		
	\item $\V_x(c, 0) = u'(c)$.
	\item If $m>0$, $\V_x(m, -) = u'(m)$, provided $m\neq R$. If $m=R$ then $V(\cdot,-)$ has left and right derivatives at $m$ and 
	
	\begin{equation}\label{disc}
	\frac{d\V(\cdot, -)}{dx}_-(m)= u'(m)\leq  \frac{d\V(\cdot, -)}{dx}_+(m).
	\end{equation}
	\end{enumerate}
\end{thm}
\begin{proof}
	\begin{enumerate}
	\item\label{one}	Fix $L<a<x<b$, and set $\t:=\t_{a} \land \t_{b}$. Since $\V(\cdot,0) \geq 0$, we obtain
	\begin{equation}\label{con1}
	\begin{split}
	\V(x,+) &\ge \E^{(x,+)}[e^{-r\t \land J }\V(S_{\tau \land J},F_{\tau \land J})] \\
	& \ge \E^x[e^{-r\t}\V(S^+_{\t},+) \1_{\t< J}]\\
	& =  \E^x\big[\E[e^{-r\t}\V(S^+_{\t},+) \1_{\t < J}|\F_{\t}]\big]\\
	& = \E^x\big[e^{-r\t}\V(S^+_{\t},+)\E[ \1_{\t < J}|\F_{\t}] \big]\\
	& = \E^x[e^{-(r+\l) \t}\V(S^+_{\t},+)]\\
	& = \V(a)\E^x[e^{-(r+\l) \t_{a}}\1_{\t_{a} < \t_{b}}] + \V(b)\E^x[e^{-(r+\l) \t_{b}}\1_{\t_{b} < \t_{a}}],
	\end{split}
	\end{equation}
	where the first inequality follows from the fact that the Snell envelope is a class D supermartingale and the second inequality from the positivity of $\V$. \\
	Recall that $\phi^f$ and $\psi^f$ denote the fundamental solutions to
	$$
	\mL^f g - rg = 0,$$	
	and let $\phi^\l$ and $\psi^\l$ denote the decreasing and increasing fundamental solutions to 
	\begin{equation}
		\mL^+ g - (r+\l) g = 0. 
	\end{equation}
	It follows, by standard arguments that, taking $a<x<b$,
	\begin{equation}
	\begin{split}
	\E^x[e^{-(r+\l) \t_{a}}\1_{\t_{a} < \t_{b}}]&=\frac{\psi^\l(b)\phi^\l(x)-\phi^\l(b)\psi^\l(x)}{\psi^\l(b)\phi^\l(a)-\phi^\l(b)\psi^\l(a)}\\
	&\text{and}\\
	\E^x[e^{-(r+\l) \t_{b}}\1_{\t_{b} < \t_{a}}]&=\frac{\psi^\l(x)\phi^\l(a)-\phi^\l(x)\psi^\l(a)}{\psi^\l(b)\phi^\l(a)-\phi^\l(b)\psi^\l(a)}.
	\end{split}
	\end{equation}
	We rewrite inequality (\ref{con1}) as
	\begin{equation}\label{con2}
	\V(x,+)\geq  \frac{\psi^\l(b)\phi^\l(x)-\phi^\l(b)\psi^\l(x)}{\psi^\l(b)\phi^\l(a)-\phi^\l(b)\psi^\l(a)}\V(a,+)+\frac{\psi^\l(x)\phi^\l(a)-\phi^\l(x)\psi^\l(a)}{\psi^\l(b)\phi^\l(a)-\phi^\l(b)\psi^\l(a)}\V(b,+).
	\end{equation}
	Dividing both sides of inequality \ref{con2} by $\phi^\l(x)$, we  deduce that $\tV:=\V(\cdot,+)/\phi^\l(\cdot)$ is $\tilde{s}$-concave on $(L,\infty)$  where  $\tilde{s}:=\psi^\l/\phi^\l$, which allows us to apply the arguments from \cite{Sam} to show \ref{one} as follows. 
	
	Standard arguments show that $\ts$ is differentiable except at $R$ and $\ts'>0$. By $\ts$-concavity $\frac{d\tV}{d\ts}_+\leq \frac{d\tV}{d\ts}_-$ and both exist. 
	
	Now, since $y\in \partial C^+$, $y\geq B>R$ and
	$$\V(y,+)={u(y)} \text{ while }\V(y-\e,+)\geq u(y-\e)\text{ and }\V(y+\e,+)\geq u(y+\e)\text{ for all small }\e>0,
	$$
	and it follows that, defining $\tu=\frac{u}{\phi^\l}$,
	$$
	\frac{d\tV}{d\ts}_+(y)\geq \frac{d\tu}{d\ts}(y)\geq	\frac{d\tV}{d\ts}_-(y)
	$$
	and so we must have equality throughout. Since $y\neq R$ we obtain the required equality.
	
	\item We proceed as in the proof of \ref{one}:
	\begin{equation}\label{con3}
	\begin{split}
	\V(x,0) &\ge \E^{x,0}[e^{-r\t}\V(S_{\tau },F_{\tau })] \\
	& \ge \E^x[e^{-r\t}\V(S^0_{\t},0)]\\
	& =  \E^x\big[\E[e^{-r\t}\V(S^0_{\t},0) |\F_{\t}]\big]\\
	& = \V(a,0)\E^x[e^{-r \t_{a}}\1_{\t_{a} < \t_{b}}] + \V(b,0)\E^x[e^{-r \t_{b}}\1_{\t_{b} < \t_{a}}].
	\end{split}
	\end{equation}
	Thus, setting $\ts^0=\frac{\psi^0}{\phi^0}$, and $\tV^0=\frac{V(\cdot,0)}{\phi^0}$, we see that $\tV^0$ is $\ts^0$-concave. Since $\ts^0$ is $C^1$ we obtain the required smoothness as above.
	\item This result follows in the same way, except there may be a problem if $m=R$, where $(\ts^-)'$ is discontinuous. Nevertheless,
	$$\V(m,-)={u(m)} \text{ while }\V(m-\e,-)= u(m-\e)\text{ and }\V(m+\e,-)> u(m+\e)\text{ for all small }\e>0,
	$$
	establishing (\ref{disc}).
	\end{enumerate}
\end{proof}

\subsection{The solution to the seller's problem}\label{sol1}
To aid our analysis, we shall henceforth assume:
\begin{assumption}
The coefficients $\sigma_f$ and $\mu_f$ are piecewise continuous on $\R_+$ for each $f\in I$, i.e. there is a finite set $\H$ such that $\sigma_f$ and $\mu_f$ are continuous on $\R_+\setminus \H$.
\end{assumption}
We will repeatedly use the corollary of the following lemma:
\begin{lemma}\label{mon}
Suppose that
$0\leq a < b <\infty$, $h:[a,b]\mapsto \R$,  and that $h$ is continuous on $[a,b]$ and has no positive (local)  maximum on the interval $(a,b)$. 
Then if either
\begin{itemize}
\item $h(a)\geq 0$ and $h$ is strictly increasing at $a$, i.e. 
$$\exists \eta>0\text{ such that for all }\epsilon\in(0,\eta),\; h(a+\epsilon)>h(a),
$$
\item[] or
\item $h(b)\geq 0$ and $h$ is strictly decreasing at $b$,  i.e. 
$$\exists \eta>0\text{ such that for all }\epsilon\in(0,\eta),\; h(b-\epsilon)>h(b),
$$
\end{itemize}
then $h$ is monotone and non-negative on $[a,b]$.
\end{lemma}
\begin{proof}
Consider the first case. Define $c:=\inf\{x\in[a,b] \text{ such that }h(x)< 0\}$, with the infimum taken as $b$ if $h$ is positive on $(a,b]$. It follows from the continuity of $h$, and the assumption that $h(a)\ge 0$,  that $h\ge 0$ on $[a,c]$.

Now define $s=\sup_{x\in[a,c]} h(x)$. Since $[a,b]$ is compact and $h$ is continuous, the supremum is attained at $\bar x$, say. By the assumption that $h$ is strictly increasing at $a$, $s>0$ and $\bar x\neq a$. Now if $\bar x< c$ then it is a positive local maximum of $h$ in $[a,b]$ which contradicts the assumption that $h$ has no positive maximum. Thus $\bar x=c$ and so $s=h(c)>0$ and this means that $c=b$ and $0\leq h(x)\leq h(b)$ for $x\in[a,b]$ and that $h$ has no local maximum on $(a,b)$. Now for any $a<b'\le b$ a similar argument shows that $s(b'):=\sup_{x\in[a,b']}=h(b')$ which immediately implies that $h$ is increasing since $b'$ is arbitrary.

The proof of the second case is similar.
\end{proof}
\begin{cor}\label{cmon}
For each $\gamma> 0$, we define 
$$
\mL^f_\gamma:g\mapsto \mL^fg-\gamma g.
$$
Then if
$\mL^f_\gamma h\geq 0$, 
in the sense that $h\in C^1(a,b)$ and is in $C^2(a,b)$ off a finite set $\H_1=\{d_1,\ldots,d_k\}$, and
either: 
\begin{itemize}
\item $h(a)\geq 0$ and $h'(a+)\geq 0$  
\item[] or
\item $h(b)\geq 0$ and $h'(b-)\leq 0$ 
\end{itemize}
then $h$ is monotone and non-negative on $[a,b]$.
\end{cor}
\begin{proof}
Recall $\phi^f_\gamma$ and $\psi^f_\gamma$ are fundamental solutions of 
$$
\mL^f_\gamma g=0
$$
with (see Remark \ref{strict}) $(\psi^f_\gamma)'>0$  and $(\phi^f_\gamma)'<0$. We set
$$
\hpsi:=\frac{\psi^f_\gamma-\psi^f_\gamma(a)}{\psi^f_\gamma(b)-\psi^f_\gamma(a)}
$$
and
$$
\hphi:=\frac{\phi^f_\gamma-\phi^f_\gamma(b)}{\phi^f_\gamma(a)-\phi^f_\gamma(b)},
$$
so that
$\hpsi(a)=\hphi(b)=0$
and
$\hpsi'>0$ and $\hphi'<0$.

Consider the first case and for any $\e>0$ define
$$
h_\e=h+\e\hpsi.
$$
By hypothesis,
$$
\mL^f_\gamma h_\e=\mL^f_\gamma h\geq 0,
$$
in the strong sense (see Ch. 9 of \cite{Gil}), so, by the strong maximum principle for elliptic operators  (see Theorem 9.6 in \cite{Gil}), $h_\e$ has no positive maximum on $(a,b)$. 

Since $h'(a)\geq 0$, $h_\e$ is strictly increasing at $a$ and $h_\e(a)=0$.
By Lemma \ref{mon}. $h_\e$ is non-negative and monotone on $(a,b)$ for any $\e>0$. Letting $\e\downarrow 0$ we see that $h$ has the same properties. The same argument, with $\hpsi$ replaced by $\hphi$, works in the second case.

\end{proof}

We are now ready to propose the candidate solution via the following free boundary problem, and we will prove the candidate solution is indeed the value function (restricted to the interval $(L,B)$ in the case $f=+$). 

Let $\tD: = \bigcup_{f \in I} \tD^f\times\{f\}$ and $\tC:= \bigcup_{f \in I} \tC^f = E\setminus\tD$, where 
\begin{equation}\label{df}
\tD^f=\begin{cases}
[0,m'],&$if $f=-\\
[B',\infty),&$if $f=+\\
[c',\infty),&$if $f=0
\end{cases}
\end{equation}
and 
\begin{equation}\label{cf}
\tC^f=\begin{cases}
(m',H),&$if $f=-\\
(L,B'),&$if $f=+\\
(L,c'),&$if $f=0.
\end{cases}
\end{equation} 

Furthermore, for $A\subset\R_+$, let $\tC_A $  denote $\tC \setminus A\times I$ and $\tC_R$ denote $\tC\setminus\{(R,-),(R,+)\}$, and $\tC_A^{\circ}$, respectively $\tC_R^{\circ}$, denote the interior of $\tC_A$, respectively $\tC_R$. 

Let $v:E \rightarrow \R$. We say $(v,B', m', c')$ is a \emph{solution to the free boundary problem} (\ref{definition: fbp}) if $v \in C(E) \cap C^1(\tC_R) \cap C^2(\tC_R^{\circ}\cap\tC_{\H}^{\circ})$ such that
\begin{equation}\label{definition: fbp}
\begin{cases}
(\mL^f - r)v(x,f) + \l\big(v(x,0) - v(x,+)\big)\1_{\{f = +\}} = 0, & $in $ \tC_R^{\circ}\cap\tC_{\H}^{\circ})\\
v(x, f) = u(x), & $in $ \tD\\
v(L, f) = v(L, -), & f \in \{+, 0\}\\
v(H, -) = v(H, +), \\
$with$\\
p_+ v_x(R+,+) = q_+v_x(R-,+), \\
p_- v_x(R+,-) = q_-v_x(R-,-), & $if $R\ge m' \\
v_x(B', +) = u'(B'),\\
v_x(m', -) = u'(m), & \text{if $m' > 0$ and $m' \ne R$}\\
v_x(c', 0) = u'(c), & \text{if  $c' \neq R$, and $c' \neq L$} \\
v(x,+) \ge u(x), &\text{on $[A, c']$ if $c'>A$}\\
A\le B', L \le c' ,$ and $ 0 \le m' < H 
\end{cases}
\end{equation}

\begin{thm}\label{thm: smoothness of V}
	The quadruplet $(\V, B, m, c)$ is a solution to the free boundary problem (\ref{definition: fbp}). 
\end{thm}
\begin{proof}
	Fix $f$. Take an open interval $I:=(y,z)$ such that $I \subset C^f\setminus(\H\cup\{R\})$. Suppose there exists a solution $v(x,f) \in C^2(I) \cap C(\bar{I})$ to the following ODE (in the case $f=+$ we implicitly assume $v(x,0)$ is already known):
	\begin{equation}\label{eq: ode}
		\begin{cases}
		(\mL^f - r)v(x,f) + \l\big(v(x,0) - v(x,+)\big)\1_{\{f = +\}} = 0, & $in $ (y,z)\\
		v(y,f) = \V(y, f), \quad v(z,f) = \V(z, f). 
		\end{cases}
	\end{equation}
	Let $\t:=\t_y\land\t_z\land\stop$ where $\stop:=\inf\{t\ge0: F_t \ne f\}$. Then, if $f\ne+$, by Dynkin's formula, we obtain
	\begin{equation}
		v(x,f) = \E^{(x,f)}[e^{-r\t} v(S_\t, F_\t)] - \E^{(x,f)}\Big[\int_0^\t e^{-rt}(\mL^f - r)v(S_t, F_t) dt\Big]. 
	\end{equation}
	We see $v(S_\t, F_\t) = \V(S_\t, F_\t)$ $P$-a.s. since $\stop\ge \t_y\land\t_z$, and $(\mL^f - r)v(S_t, F_t) = 0 $ on $[0, \t)$. This leads to 
	\begin{equation}
		v(x,f) = \E^{(x,f)}[e^{-r\t} \V(S_\t, F_\t)] = \V(x,f),
	\end{equation}
	where the second inequality holds since $e^{-rt\land \t} \V(S_{t\land \t}, F_{t\land \t})$ is a martingale. If $f = +$, we need to define $h(x,l):= v(x,l)\1_{l = +} + \V(x,0)\1_{l=0}$. For any $\e>0$, we can still apply Dynkin's formula for the stopping time $\t_\e := (\t-\e)^+$ and get
	\begin{equation*}
	h(x) = \E^{(x,f)}[e^{-r\t_\e} h(S_{\t_\e} , F_{\t_\e} )] - \E^{(x,f)}\Big[\int_0^{\t_\e} e^{-rt}\Big\{(\mL^f - r)v(S_t, F_t)+\l\big(v(S_t,0) - v(S_t,+)\big)\Big\} dt\Big]. 
	\end{equation*}
	Since the integrand of the $dt$ term is $0$ and $h$ is bounded on $I$, we can apply dominated convergence and take $\e$ to $0$ to show
	\begin{equation}
		v(x,f) = h(x) = \E^{(x,f)}[e^{-r\t} h(S_{\t} , F_{\t} )] = \E^{(x,f)}[e^{-r\t} \V(S_{\t} , F_{\t} )] = \V(x,f). 
	\end{equation}
	The existence of a classical solution to ODE (\ref{eq: ode}) in the case $f \ne +$ simply follows from  Theorem 6.2.4 in \cite{Fri}. If $f = +$, since we have shown $v(x,0) \in C^2(I)$ and $v(x,0) = \V(x,0)$, the result in \cite{Fri} can still be applied. \\
	Next, if we can show that 
	$$\V_x(d+,f) = \V_x(d-,f) \text{ for }d\in\H\cap C^f,
	$$
	and
	$$p_f\V_x(R+,f) = q_f\V_x(R-,f)\text{ if }R\in C^f,
	$$
	then the proof is completed as the rest of assertions in (\ref{definition: fbp}) are straightforward to verify. By now we know $v(x,f)$ is a piecewise $C^2$ function for fixed $f$, and hence it can be written as the difference of two convex functions. Thus, we can apply the symmetric Ito-Tanaka-Meyer  formula, which shows, on $\{t\le \t_D\}$,
	\begin{IEEEeqnarray*}{rl}
	d e^{-rt}\V(S_t, F_t) = e^{-rt}&\Biggl[ (\mL - r)\V(S_t, F_t) \1_{S_t \not\in \{R\}\cup\H} dt+(p_{F_{t}}\V_x(R+,F_t)-q_{F_{t}}\V_x(R-,F_t)) dl^R_t \Biggr]\\
	+e^{-rt}& \half\sum_{d\in \H}(\V_x(d+,F_t)-\V_x(d-,F_t))dl^d_t+dM_t\\
	= e^{-rt}&\biggl[(p_{f}\V_x(R+,F_t)-q_{F_t}\V_x(R-,F_t)) dl^R_t +\sum_{d\in\H}(\V_x(d+,F_t)-\V_x(d-,F_t))dl^d_t\biggr] +dM_t,
	\end{IEEEeqnarray*}
	where $M_{t\land \t_D}$ is a uniformly integrable martingale. Since $e^{-rt\land \t_D}\V(S_{t\land \t_D}, F_{t\land \t_D})$ is a martingale, it follows that $\int [p_{f}\V_x(R+,f)-q_{f}\V_x(R-,f)]dl^R_t = 0$ if $R\in C^f$ and $\int\V_x(d+,f)-\V_x(d-,f))dl^d_t=0$, for each $d\in\H\cap C^f$, for Lebesgue a.a. $t$, and hence that
	$$p_{f}\V_x(R+,f)-q_{f}\V_x(R-,f)=0 \text{ if }R\in C^f\text{ and }\V_x(d+,f)-\V_x(d-,f))=0\text{ if }d\in C^f.
	$$
	
\end{proof}

\begin{thm}\label{fbpsoln}
		$\V$ is the maximal solution of  (\ref{definition: fbp}). If  $c\leq B$, then $\V$ is the {\em unique} solution and $\tD = \D$. Moreover, if $(v,B',m',c')$ is a solution with $c'\leq B'$ then $v=\V$ and it is the {\em unique} solution to (\ref{definition: fbp}).
\end{thm}

\begin{proof}
First, extend the definition of $v$ to $E$ by setting 
$$
v(x,f)=u(x)\text{ for }\ (x,f)\in\tD.
$$

Define $N_t := e^{-rt}v(S_{t},F_{t})$. As noted in Remark \ref{P2}, to show that $v\leq \V$, it is sufficient to prove $N$ satisfies property P2  from Lemma \ref{P123}.
If we can show, in addition, that $v$ satisfies properties P1 and P3 then $v=\V$.


(P2) Since $v(x, f)$ is a continuous function for each fixed $f$, there is a constant $M$ such that $v(x, f) \leq  M$ if $x \leq B'$. 
So,
\begin{equation}\label{solutionE1}
|N_{t}|=e^{-r{t}}|v(S_{t},F_{t})|\le e^{-rt}(|M| \vee |u(S_{t})|) \le |M| \vee e^{-rt}|u(S_{t})|.
\end{equation}
Hence,
\begin{equation}
\E^{(x,f)}[\sup_{\t} |N_{\t}|] \le \E^{(x,f)}[|M| \vee \sup_{\t} e^{-r\t}|u(S_{\t})|] \le |M| + \E^{(x,f)}[ \sup_{\t} e^{-r\t}|u(S_{\t})|] < \infty,
\end{equation}
by Assumption \ref{SA2}, 
which implies that $N_{t}$ is of class D.\\

Consider the sequence of stopping times $J_n$, where $J_n$ is the $n^{th}$ time that $F$ jumps from $+$ to $0$ and $J_0 = 0$. Then, defining
\begin{equation}
	A_t := \sum_{n=0}^{\infty} \1_{t \ge J_n}\triangle N_{J_n},
\end{equation}

\begin{equation}\label{eq: compensator}
	N_t = N_0 +  \sum_{n = 0}^{\infty} \int \1_{t\in [J_n, J_{n+1})} dN_t + A_t. 
\end{equation}
Since $|v(x,+) - v(x,0)| \le \sup_{x \in [L, B'\vee c']} \{v(x,+) +v(x,0)\} \leq 2M$ for all $x \ge L$, we see that the jumps of $A_t$ can be bounded by some constant denoted by $k$ . Therefore the variation process $|A_t|$ is bounded by $k\Lambda_t$ where $\Lambda_t$ denotes a Poisson process with intensity $\l$. Let $A^0$ be the compensator of $A$ . It can be shown easily that
\begin{equation}
	dA^0_t = e^{-rt}\l(v(S_t,0) - v(S_t,+)) \1_{\{F_t=+\}}dt. 
\end{equation}
Adding and subtracting $A^0_t$ in equality (\ref{eq: compensator}) and applying the symmetric Ito-Tanaka-Meyer  formula, it is evident that
\begin{equation}\label{equation: change of variable}
\begin{split}
dN_t  =
& e^{-rt}\Big[\big(-rv(S_t,+)+\mL^+v(S_t,+) + \l\big[v(S_t, 0)-v(S_t, +)]\big)\mathbbm{1}_{\{F_t=+, S_t \ne B', S_t \ne R\}}dt \\
&+\big(-rv(S_t,0)+\mL^0v(S_t,0)\big)\mathbbm{1}_{\{F_t=0, S_t \ne c'\}}dt\\
&+\big(-rv(S_t,-)+\mL^-v(S_t,-)\big)\mathbbm{1}_{\{F_t=-, S_t \ne m', S_t \ne R\}}dt\\
&+\big(p_{F_t}v_x(R+,F_t) - q_{F_t}v_x(R-,F_t)\big)  \1_{F_t \ne 0}dl^R_t +(v_x(B'+, +) - v_x(B'-, +)) \1_{F_t=+}dl^{B'}_t\\
&+(v_x(c'+, 0) - v_x(c'-, 0)) \1_{\{F_t=+, c'>L\}}dl^{c'}_t \\
& +(v_x(m'+, 0) - v_x(m'-, 0)) \1_{\{F_t=+, m' \ne 0\}}dl^{c'}_t\Big] + dM_t, 
\end{split}
\end{equation}
where $M_t$ is a local martingale.

By setting $\t := \t_\tD$, it is not hard to check from (\ref{equation: change of variable}) that $N_{t\land\t}$ is a martingale. Hence P2 holds and $v\leq \V$.

(P1) By (\ref{definition: fbp}) and Assumption \ref{c}, all local time terms in (\ref{equation: change of variable}) vanish and the $dt$ terms are all non-positive. Thus, $N_t$ is a class D supermartingale. \\

(P3) First, we show $v(x, -) \ge u(x)$ on $[m', H)$.\\
Set $g:=v(\cdot,-)-u$.

First, suppose that $m' \ge R$, then, on $(m',H)$
\begin{displaymath}
\mL^-g-rg  = \mL^-v-rv -(\mL^-u-ru)=-(\mL^-u-ru) \ge 0,
\end{displaymath}
by Assumption \ref{c}.
Note that, if $m'>R$, $g(m')=g'(m')$ by smooth pasting  while, in the case $m' = R$, $g(R)=0$ and $v_x(R+,-) \ge u'(R)$ so that $g'(R)\geq 0$. Therefore,  by Corollary \ref{cmon}, $g$ is monotone increasing on $[m',H)$ and so $v(x,-) \ge u(x)$.\\

Next, suppose that $m'<R$. On $(m', H) \setminus \{R\}$, we still have 
\begin{equation}
	\mL^- g - rg = ru - \mL^-u > 0.
\end{equation}
Applying Corollary \ref{cmon} to $g$ on the interval $[m',R]$ we conclude that $g$ is monotone increasing on the interval.

Moreover, from the monotonicity of $g$, we see that $v_x(R-, -) \ge u'(R) \ge 0$. Therefore,
\begin{equation}
	g'(R+) = \frac{q_-}{p_-}v_x(R-, -) - u'(R) \geq v_x(R-, -) \ge u'(R)\geq 0, 
\end{equation}
while $g(R)>0$.
Thus, we can apply the corollary once more to see that $g$ is non-negative on $[m',H]$. \\

The proof that $v(x,0) \ge u(x)$ follows exactly the same lines.\\

Finally, to prove $v(x, +) \ge u(x)$, now set $g(x):=v(x,+) - u(x)$ and $\gamma=r+\l$.
Recall that $\psig$ and $\phig$ denote the increasing and decreasing fundamental solutions respectively to the ODE:
\begin{equation}
\mL^+ w(x) - \gamma w(x) =0. 
\end{equation}
We divide the proof into two cases:\\
\textit{Case 1}. $c'\leq A$. Note that on $(A,B')$, by Assumption \ref{c} 
	\begin{IEEEeqnarray}{rl}
		\mL^+ g(x) - \gamma g(x) &=\mL^+v(x,+)-(r+\l)v(x,+)-(\mL^+u(x)-ru(x))+\l u(x)\\
		&\ge \l u(x) -\l v(x,0) = 0,
	\end{IEEEeqnarray}
and, by smooth pasting, $g(B')=g'(B')=0$. Thus, by Corollary \ref{cmon}, $g\geq 0$ on $[A,B']$ and so $v(\cdot,+)\geq u$ on $[A,B']$. \\
Conversely, by Assumption \ref{c} and the fact, established above, that $v(x,0) \ge u(x)$, we have
\begin{equation}\label{eq: v-u}
\mL^+ g(x) - (r+\l) g(x) \le \l u(x) -\l v(x,0) \le 0.
\end{equation}
Suppose that $g(R) \ge 0$. Then, since $g(A) \ge 0$ and $g(L) = v(L,-) - u(L)\ge0$, we can use the strong {\em minimum} principle to deduce that $g\ge0$ on $[L,A]$. 

Now suppose, to the contrary, that  $g(R)<0$, while $g(L),g(A)>0$. Denote the last zero of $g$ on $[L,R)$ by $Z$ then, by the strong minimum principle (see Theorem 9.6 in \cite{Gil}),  we must have $g$ decreasing on $(Z,R)$  and so $g'(R-) \le 0$. Conversely, denoting the first zero of $g$ on $(R,A)$ by $Z'$, $g$ must be increasing on $[R,Z]$ and so $g'(R+) \ge 0$. However, this implies that
\begin{equation}\label{ineq}
0\le p_+g'(R+) - q_+g'(R-).
\end{equation}
Now, \ref{definition: fbp} tells us that 
\begin{equation}
p_+g'(R+) - q_+g'(R-)= (q_+-p_+)u'(R).
\end{equation} 
But
$$(q_+-p_+)u'(R)< 0,
$$
since, by assumption, $p_+>q_+$ and $u'>0$ by assumption.
This gives a contradiction and we conclude that $g(R)\geq 0$ as required. \\

\textit{Case 2}. $B'\geq c' >  A$. We can  prove that $g\ge0$ on $[c',B']$ and on $(L,A]$ by following the same argument as in Case 1. The free-boundary problem {\em posits} that $g\ge 0$ on $[A,c']$. 

It follows that $v$ satisfies property P3 and hence equals $\V$.
\end{proof}
\begin{remark}
We underline that, in order to prove uniqueness of the solution to the free boundary problem we have had to assume that $c\leq B$ and to require that the solution dominates $u$ for $(x,f)\in (A,c')\times\{ +\}$.
\end{remark} 
\subsection{An example and a counterexample}\label{example seller}
Here we present two examples where we calculate as much as we can do in closed form. The first will be an example where all assumptions are satisfied. The second is a counterexample where we demonstrate what can happen if Assumption \ref{c} does not hold. 
Section \ref{risk aversion} will present further numerical examples.\par

\begin{example}\label{ex}
Let $u(x)=x^\half$. Set 
$r=\frac{3}{4}$, 
$\l=\frac{1}{4}$, $L=\half$, 
$R=\frac{10}{13}$, $\frac{p_+}{q_+}=1.63$, $\frac{q_-}{p_-}=9.6$, 
and 
$H=\frac{13}{10}$.
Then we take 
$\mm(x)=\frac{17}{12}x$, 
$\sm^2(x)= \frac{5}{6}x^2$, 
$\m_+(x)= (x+1)$, 
$\sigma_+^2(x)= 2x^2$, 
$\mo(x)=x$, and 
$\so^2(x)=\frac{10}{3}x^2$.
It is not hard to see that Assumptions  \ref{c} and (thanks to Theorem \ref{suff} and Assumption \ref{suffa}) 
\ref{assumption: utility} both hold and that, since
$\mL^+u(x)=\half(x^{-\half}-x^\half)$, $A=1$.  

As we shall see, $v_0=u$, and $m>L$.
Thus, we can find the value function by solving the free boundary problem. 
The ODE
\begin{displaymath}
\mL^-v-rv=\frac{5}{12} x^2v''(x)+\frac{17}{12}xv'(x)-\frac{3}{4}v(x)=0,
\end{displaymath}
admits a general solution of the form $v(x,-)=E x^{\frac{3}{5}}+Fx^{-3}$. 

The ODE
\begin{displaymath}
\mL^+v-rv-\l(v-u)=x^2v''(x)+(x+1)v'(x)-v(x)+\frac{1}{4}x^\half=0 
\end{displaymath}
has general solution $v(x,+)=Cxe^{\frac{2}{x}}+D(x+1)+\frac{\sqrt{\pi}}{6}xe^{\frac{1}{x}}\erf(x^{-\half})$, where $\erf$ is the standard error function, 
$$\erf:x\mapsto \frac{2}{\sqrt{\pi}}\int_0^x e^{-t^2}dt.
$$

Assuming that $m' \ge L$, we compute the value of $B'$ by smooth pasting and by using  the boundary conditions
$$
v(L+,+)=v(L,-)=u(L),\; v(R-,+)=v(R+,+)\text{ and }p_+v'(R+,+)=q_+v'(R-,+).
$$ 
Numerical approximation gives 
$B'=2.002918$ and 
$$v(x,+)=-0.200733xe^{\frac{2}{x}}+0.470372(x+1)+\frac{\sqrt{\pi}}{6}xe^{\frac{1}{x}}\erf(x^{-\half})
$$
for $x\geq R$, 
and  
$$
v(x,+)= -0.430140xe^{\frac{2}{x}}+0.836356(x+1)+\frac{\sqrt{\pi}}{6}xe^{\frac{1}{x}}\erf(x^{-\half}),
$$
for $x\leq R$.

Then by smooth pasting at $m'$ and the condition $v(H-,-)=v(H,+)$, we compute $m'=0.8$ and $$v(x,-)=\frac{\sqrt{5}}{9}((\frac{1}{10}({\frac{5x}{4}})^{-3}+\frac{7}{2}({\frac{5x}{4}})^{\frac{3}{5}}).
$$

Thus, by Theorem \ref{fbpsoln}, since $c'=L<B'$, we derive the value function $\V(x,f)$:
\begin{equation}
\V(x,f) = \left\{ \begin{array}{ll}
-0.430140xe^{\frac{2}{x}}+0.836356(x+1)+\frac{\sqrt{\pi}}{6}xe^{\frac{1}{x}}\erf(x^{-\half}) & \textrm{if $x \in (\half,\frac{10}{13})$, $f=+$}\\
-0.200733xe^{\frac{2}{x}}+0.470372(x+1)+\frac{\sqrt{\pi}}{6}xe^{\frac{1}{x}}\erf(x^{-\half})&\textrm{if  $x\in(\frac{10}{13},2.00292)$, $f=+$}\\
\frac{\sqrt{5}}{9}((\frac{1}{10}({\frac{5x}{4}})^{-3}+\frac{7}{2}({\frac{5x}{4}})^{\frac{3}{5}})& \textrm{if $x \in (0.8,\frac{13}{10})$, $f=-$}\\
x^{\half} & \textrm{otherwise.}\\
\end{array} \right.
\end{equation}
The optimal strategy is to sell the stock when its price is higher than $2.002918$ in the positive regime or lower than $0.8$ in the negative regime and to sell everywhere in the 0 regime.

\end{example}

\begin{example}
Let $u(x)=80(0.1)^{\frac{1}{4}}+5(0.1)^{-\half}x^{\frac{3}{4}}$. 
Set 
$r=\frac{7}{8}$, $\l=0$, $L=0.5$, $R=\frac{10}{13}$, $\frac{p_+}{q_+}=1$, 
$\frac{q_-}{p_-}=1$, and 
$H= 0.7$. 
Then we take 
$\mm(x)=0$, $\sm^2(x)= x^2$, $\m_+(x)= 2x$, 
$\sigma_+^2(x)= 
x^2$ for $x\leq 9.6$ and $\sigma_+^2(x)= 8x^2$ for $x>9.6$, 
$\mo(x)=0$
and $\so^2(x)=x^2$.

It follows that $\mL^-u<0$ and $\mL^0u<0$, but $\mL^+u(x)>0$ iff $x\in (7.84222,9.6)$ contradicting Assumption \ref{c}.
As we shall see, $v_0=u$, and $C=(t,s)$ for some $s>9.6>7.84222>t>H$. Thus, we can find the value function by solving the free boundary problem:\par

\begin{displaymath}
\mL^+v-rv=0 
\end{displaymath}
with smooth pasting at $t$, $s$ and $9.6$.

The ODE 
$$
\half x^2f''+2xf'-\frac{7}{8}f=0
$$
has general solution $v(x,+)=Ax^{-\frac{7}{2}}+Cx^{\frac{1}{2}}$, 
while the ODE
$$
4x^2f''+2xf'-\frac{7}{8}f=0
$$
has general solution $v(x,+)=Dx^{-\frac{1}{8}}+Ex^{\frac{7}{8}}$. The smooth pasting condition and numerical  approximation gives 
$t=6.875304$, $s=10.174128$ and 

$$v(x,+)=\begin{cases}
u(x)&x<t\\
1216.52x^{-\frac{7}{2}}+42.2158x^{\frac{1}{2}}&t\leq x\leq 9.6\\
67.6526x^{-\frac{1}{8}}+11.0910x^{\frac{7}{8}}&9.6<x\leq s\\
u(x)&x>s.
\end{cases}$$

The optimal strategy is to sell the stock when its price is lower than $t=6.87530$ or greater than $s=10.1741$ in the positive regime and everywhere in the negative regime and everywhere in the 0 regime. Note that in this example, the zero regime can only be reached by starting there. 
Note also that smooth pasting was required at 9.6 due to the discontinuity of the diffusion coefficient.  

\end{example}

\section{The Buyer's Problem}\label{buyer's problem}
\subsection{The problem}
If traders want to find the best time to purchase a stock and sell it later to maximise their {\em incremental} expected utility, they will try to solve the following double optimal stopping problem:
\begin{equation}\label{prob2}
\V_p(x,f):=\sup_{\tau_1 < \t_2}\E^{(x,f)}[e^{-r\t_2}u(S_{\t_2}) - e^{-r\t_1}u(S_{\t_1})].
\end{equation}
In other words, we would like to maximize the marginal utility of buying and selling, where $\t_1$ (resp. $\t_2$) is interpreted as the buying (resp. selling) time. We call this the  {\em buyer's problem}. By Lemma \ref{appendix: equivalence}, the buyer's problem admits an equivalent formulation given by

\begin{equation}\label{BP}
\V_p(x,f)=\W(x,f):=\sup_{\tau}\E^{(x,f)}[e^{-r\tau}g(S_\t, F_\t)],
\end{equation}
where $g(x,f) := \V(x,f) - u(x)$ and $\V$ is the value function of the seller's problem (\ref{SP}). We assume $u$ satisfies  Assumption \ref{assumption: utility} and \ref{c}. This implies 
\begin{equation}\label{assumption: sign of g}
\begin{cases}
\mL^- g-rg > 0, & \text{in $(m,H)\setminus\{R\}$}\\
\mL^+ g-rg < 0, & \text{in $(L,A)\setminus\{R\}$}\\
\mL^+ g-rg > 0, & \text{in $(A,B)$}.
\end{cases}
\end{equation}

\begin{thm}
The gains function, $g$ is continuous,  positive, and bounded. It follows that the value function, $\W$, is also. 
\end{thm}
\begin{proof}
We have established that $\V$ is continuous and $u$ is continuous by assumption so $g$ is continuous.
Positivity follows from the fact that $\V\geq u$. 
Now $C^0$ and $C^-$ are both bounded so to establish that $g$ is bounded we need only consider its behaviour on $C^+$. Recall from Theorem \ref{theorem: D shape} that $C^+$ is bounded unless $m<L$,  $B<c$ and $\l>0$. In that case, equation (\ref{v-u2}) shows that, for $z\ge c$,
$$
0\le g(z,+)=\V(z,+)-u(z)\le \frac{\phi^+_{\l+r}(z)}{\phi^+_{\l+r}(c)}(\V(c,+)-u(c)),
$$
since $\mL^+u-ru<0$ on $(c,\infty)$ if $c>A$ (by Assumption \ref{c}) and $B\ge A$.
Now we know that $\lim_{z\rightarrow\infty}\phi^+_{\l+r}(z)=0$ so $\lim_{z\rightarrow \infty}g(z,+)=0$, and so,  since $g$ is continuous, we may conclude that  $g$ is bounded.
\end{proof}

There are a few direct consequences based on the formulation.
Denote the stopping set for (\ref{BP}) by $\hD$ and the continuation set by $\hC$ .We conclude that $\t_\hD$ is an optimal Markov time by Shiryaev \cite{Shir} Chapter 3 Theorem 3, and $e^{-rt}\W(S_t,F_t)$ is the Snell envelope of  $e^{-rt}g(S_t, F_t)$ and is a class D positive supermartingale. 

Further define $\hD^f=\{x:\; (x,f)\in \E\text{ and }\W(x,f)=g(x,f)\}$ and $\hC^f\{x:\; (x,f)\in \E\text{ and }\W(x,f)>g(x,f)\}$ for $f \in \I$. Then
 \begin{thm}[]~\label{buyer shape}
	\begin{enumerate}
				\item   $\hD^-=\emptyset$ if $0$ is inaccessible or $\hD^-=\{(0,-)\}$ if $0$ is absorbing.
				\item 	$\hD^0=\emptyset$.
				\item 	$\hD^+\subseteq (L, A]$ and if $\l=0$ then $\hD^+$ is of the form $[a,b]$ with $L<a<b\leq A$.
	\end{enumerate}
\end{thm}
\begin{proof}

\begin{enumerate}
\item\label{minus} It is sufficient to show that $\hC^- = (0,H)$. 

If $x \in (0,m]$, then $g(x,-)=0$ which is the global minimum of $g$. Since $(S,F)$ is regular except at $(0,-)$ and there are points  $(y,f)\in \E$ with $g(y,f)>0$ it follows that $(x,-)\in \hC^-$.

For $x\in (m,H)$, define $\t:=\t^-_{m}\land\t^-_H$. Then by (\ref{c}) and the Ito-Tanaka-Meyer  formula,

\begin{equation}
\begin{split}
	\E^{x,-}[e^{-r\t}g(S_\t, -)] & = g(x,-)+\E^{x,-}[\int_{0}^{\t} e^{-rt}\{\mL^--r\}g(S_t, -) \1_{S_t \ne R}dt] \\
	&+ (p_-g_x(R+,-) - q_-g_x(R-,-)) \mathbb{E}^{x,-}[\int_0^\t e^{-rt} dl^R_t] \\
	& = g(x,-)+\E^{x,-}[\int_{0}^{\t} e^{-rt}\{\mL^- -r\}g(S_t, F_t) \1_{S_t \ne R}dt]\\
	& +(q_- - p_-)u'(R)\mathbb{E}^{x,-}[\int_0^\t e^{-rt} dl^R_t] \\
	&> g(x,-), 
\end{split}
\end{equation}
the last inequality following from the fact that $q_->p_-$, $u$ is increasing and 
$$
\mL^-g -rg=ru-\mL^-u>0\text{ on }(m,H)\setminus \{R\}.
$$
Thus, $R\in\hC^-$ and $\hC^- = (0,H)$.

\item Recall that, by Assumption \ref{c}, $\mL^0u-ru<0$ on $(L,\infty)$.
 Consider $x\in (L,\infty)$. \\
 Suppose that $x\in\tD^0$. Then $g(x,0)=0$ and we can argue as in \ref{minus} that $x\in\hC^0$.
 
 Conversely, suppose that $x\in\tC^0$, in which case $\mL^0g(\cdot,0)$ is positive in an interval $(y,z)\subset (L,\infty)$ containing $x$. It is sufficient to exhibit a Markov time $\t$ with $\E^{x,-}[e^{-r\t}g(S_\t, 0)]>g(x,0)$.

Setting $\t:=\t^0_{y}\land\t^0_z$, we see that
\begin{equation}
\begin{split}
	\E^{x,0}[e^{-r\t}g(S_\t, F_\t)] & = g(x,0)+\E^{x,0}[\int_{0}^{\t} e^{-rt}\{\mL^0-r\}g(S_t, 0) dt] \\
	&> g(x,0), 
\end{split}
\end{equation}
the inequality following from the fact that 
$$
\mL^0g -rg=ru-\l^-u>0\text{ on }(L,\infty).
$$

\item We can show $\hD^+ \cap(A, \infty) = \emptyset$ in exactly the same fashion. 
Moreover,  since $L\not\in \hat D^-$, we can see that, by continuity, $\lim_{x \downarrow L}\W(x,+) = \W(L,-) > g(L,-) = g(L+,+)$, which ensures $\inf \hD^+ >L$. 

To complete the proof when $\l=0$, 
suppose $\hD^+$ is not connected. Then there exists an interval $(y,z) \subset \hC^+$ such that $y,z \in \hD^+$. For $y \in (y,z)$, it is optimal to stop at $\t:= \t_{y} \land \t_{z}$. Therefore, by (\ref{assumption: sign of g}), 
\begin{equation}
\begin{split}
\W(y,+) & = \E^{y,+}[e^{-r\t}g(S_\t, +)]\\
& = g(y,+)+\E^{y,+}[\int_{0}^{\t} e^{-rt}\{\mL^+-r\}g(S_t, F_t) \1_{S_t \ne R}dt]+(q_+ - p_+)u(R)\mathbb{E}^{y,+}[\int_0^\t e^{-rt} dl^R_t] \\
&< g(y,+),
\end{split}
\end{equation}
a contradiction. Therefore, $\hD^+$ is an interval. 
\end{enumerate}
\end{proof}

The smooth pasting conditions still hold at the boundary of $\hC^+$: 
\begin{thm}[]~
	$\W_x(z, +) = g_x(z,+)$for any  $z\in\partial\hC^+\setminus \{R\}$.  
\end{thm}
\begin{proof}

Since $g\geq 0$, $\W\geq 0$ and so we may argue as in the proof of Theorem \ref{theorem: smoothpasting} that $\W(\cdot,+)$ is $\ts$ concave.

Now take $z\in\partial\hC^+\setminus \{R\}$ then
	$$\W(z,+)={g(z,+)} \text{ while }\W(z-\e,+)\geq g(z-\e,+)\text{ and }\V(z+\e,+)\geq g(x+\e,+)\text{ for all small }\e>0,
	$$
	and it follows that, defining $\tg:=\frac{g}{\phi^\l}$ and $\tW:=\frac{\W}{\phi^\l}$,
	$$
	\frac{d\tW}{d\ts}_+(z)\geq \frac{d\tg}{d\ts}(z)\geq	\frac{d\tW}{d\ts}_-(z)
	$$
	and so we must have equality throughout. Since $z\neq R$ we obtain the required equality.
\end{proof}

Let $w:E \rightarrow \R$. We call $(w,\tC)$ a solution to the free boundary problem if $w$ is non-negative, $\tC=\tC^+\times \{+\}\cup(L,\infty)\times \{0\}\cup (0,H)\times \{-\}$ where $\tD:=\E\setminus\tC\subseteq (L,A]$, $\tD^+:=(L,\infty)\setminus\tC^+$ is non-empty and closed;
and
$w \in C(E) \cap C^1(\tC_R\cup \tD)$ with $w(\cdot,+)\in C^2(\tC_R^{\circ})$ such that
\begin{equation}\label{definition: buyer fpb}
\begin{cases}
(\mL^f - r)w(x,f) = 0, & $in $ \tC_R^{\circ}\\
w(x, f) = g(x), & \text{in } \tD\\
w(L+, +) = w(L+,0)=w(L, -), \quad \lim_{x \to \infty }w(x,+) = 0,\\
w(H-, -) = w(H, +), \quad w(0,-) = 0,\\
p_f w_x(R+,f) = q_fw_x(R-,f), & f\in \{-,+\} \text{ and }  (R, f) \notin \tD^\circ \\
w_x(z, +) = g_x(z,+), & \text{if }z\in\partial \tC^\setminus\{R\}\\
w(x, f) \ge 0, &\text{in $E$}
\end{cases}
\end{equation}

\begin{thm}\label{thm: smoothness of W}
	The triplet $(\W, \hC)$ is a solution to the free boundary problem (\ref{definition: buyer fpb}) and $\W$ is maximal among such solutions. 
	If $\l=0$ then $(\W, \hC)$ is the unique solution.
\end{thm}
\begin{proof}
	The smoothness of $\W$ away from $R$ can be proved via similar arguments to those in the proof of Theorem \ref{thm: smoothness of V}. Boundary conditions are easy to show. Finally, using the symmetric Ito-Tanaka-Meyer  formula, we can argue as in the proof of Theorem \ref{thm: smoothness of V} to show $p_f \W_x(R+,f) = q_f\W_x(R-,f)$ if $R \in \hC^f$. 

To show maximality, it is sufficient (by Lemma \ref{P123}) to show that if $(w,\tD)$ is a solution to  (\ref{definition: buyer fpb}) then $w$ has properties P1-P2. 
This can be proved in exactly the same way as Theorem \ref{fbpsoln}.

To show that $w$ has property P3 under the assumption that $\l=0$, let $h(x):=w(x,+) - g(x,+)$. We have to prove $h\ge0$ on $(L,\infty)$. Define
$$\a:=\inf \tD^+\text{ and }\b:=\sup\tD^+.
$$

Let us first assume that $R\ge \b$.
Observe that $L^+_rh=L^+_rw-L^+_r\V+L^+_ru=L^+_ru\geq 0$ on $(L,\a)$ while $h(\a)=0$.
So by Corollary \ref{cmon}, $h(x) \ge 0$ on $(L,a)$. 

Moreover,  $\mL^+h-rh\ge 0$ on $(\b,R) \cup (R,A)$. Since $h(\b)= 0$ and $h'(\b) \geq 0$, Corollary \ref{cmon} tells us that $h(x)\ge0$ for all $x \in (\b,R)$ and $h'(R-)\ge0$. Furthermore, as 
\begin{equation}
p_+h'(R+)-q_+h'(R-) = (p_+-q_+)u'(R) >0,
\end{equation}
we have $h'(R+)\geq 0$. Combining this with the fact that $h$ is continuous at $R$, by Corollary \ref{cmon} $h$ is increasing on $[\b,A]$, which implies $h(x)\ge 0$ on $[\b,A]$.

Similarly, if $(y,z)$ is a maximal open sub-interval of $\tC^+\cap(\a,\b)$ we see that $\mL^+h-rh\ge 0$ on $(y,R\wedge z) \cup (R\vee y,z)$ and $h(y)=h(z)=h'(y)=h'(z)=0$ and the same argument shows that $h\geq 0$ on $[y,z]$. Thus, $h\geq 0$ on $[\a,\b]$.

Now we know $h(A) \ge 0$. {\em Assume} that  $h(B) = 0$. Since $\mL^+h-rh \le 0$ on $[A,B]$, the strong minimum principle implies $h \ge 0$ on $[A, B]$. 

To show that $h(B)>0$, first observe that $\b<B$ and so $\V(\b)>u(\b)$ and so $g(\b)>0$. 
Now if $(S_0,F_0)=(B,+)$ then $e^{-r(t\wedge {\tau^+_{\b}})}w(S_{t\wedge {\tau^+_{\b}}})$ is a bounded martingale by virtue of 
(\ref{definition: buyer fpb}). Then applying the Optional Sampling Theorem and the fact that $\b\in(L,B)$ $$
w(B)=\E^{B,+}e^{-r{\tau^+_{\b}}}w(\b)=\E^{B,+}e^{-r{\tau^+_{\b}}}g(\b)>0,
$$
so that $w(B)>0$. 
Conversely, $g(B)=\V(B)-u(B)=0$ so that $h(B)>0$. 
Finally, $\b<B$ so that $h>0$ on $[B,\infty)$.

Next, let us show $w(x,-) \ge g(x, -)$. Set $h(x):=w(x,-) - g(x,-)$. We know $h(x) = 0$ on $[0,m]$. Moreover, by the boundary condition, $h(H) = w(H,+)-g(H,+)\ge0$. Note $\mL^- h - rh <0$ on $(m, H) \setminus\{R\}$. Thus, if $R\le m$, by the strong minimum principle, we conclude $h\ge0$ on $[m,H]$. \\
Now assume $R> m$. Notice that $h(m)\ge0$, $h_\e(H)\ge0$, and $\mL^- h - rh <0$ on $(m, H)\setminus\{R\}$. If $h(R)\ge0$, by the strong minimum principle, we get $h\ge0$ on $[m,H]$. 

So, let us suppose, to the contrary,  that $h_\e(R) < 0$. By the strong minimum principle, it is necessary that $h_\e'(R-) \le 0$ and $h_\e'(R+) \ge 0$ because otherwise there would be a negative minimum on $(m, R)$ or $(R, H)$. From the smoothness conditions at $R$, it follows that
\begin{equation}
q_-h_\e(R -) - p_-h_\e(R_+) = (q_- - p_-)u'(R) > 0,
\end{equation}
which implies $h_\e'(R -) > h_\e'(R_+)$. This leads to a contradiction. 
\end{proof}

\subsection{Example \ref{ex} revisited}\label{example buyer}
\begin{example}\label{ex sell}
Recall Example \ref{ex}. We now solve the purchase problem. 

The ODE
\begin{displaymath}
\mL^0v-rv=\frac{5}{3} x^2v''(x)+xv'(x)-\frac{3}{4}v(x)=0,
\end{displaymath}
admits a general solution of the form $v(x,0)=C x^{\frac{9}{10}}+Dx^{-\frac{1}{2}}$. 
Since $w_0$ is bounded we must have
$w_0(x)=Dx^{-\frac{1}{2}}$ for a suitable positive $D$.

The ODE
\begin{displaymath}
\mL^+v-rv-\l(v-u)=x^2v''(x)+(x+1)v'(x)-v(x)+Dx^{-\half}=0 
\end{displaymath}
has general solution $v(x,+)=Axe^{\frac{1}{x}}+B(x+1)+
k(x)$ with $k:x\mapsto \sqrt{\pi}xe^{\frac{1}{x}}\erf\left(\frac{1}{x^{\frac{1}{2}}}\right)-2x^{\frac{1}{2}}$.

We guess that $\hD^+=[\a,\b]$. Continuity, smooth pasting and the boundary conditions then give $\a=0.703789$ and $\b=R=10/13$ and the value function  given below: 
\begin{equation}
\V_p(x,f) = \left\{ \begin{array}{ll}
-0.0802030xe^{\frac{1}{x}}+0.199138(x+1)\\
\phantom{000000} + 0.00291730k(x)& \textrm{if $x \in (L,\a)$, $f=+$}\\
g(x,+)=\V(x,+)-u(x)& \textrm{if $x \in [\a,R]$, $f=+$}\\
0.119001(xe^{\frac{1}{x}}-(x+1))+0.00291730k(x)& \textrm{if $x \in (R,\infty)$, $f=+$}\\
0.0250134x^{\frac{3}{5}} &\textrm{if  $x\in(0,R)$, $f=-$}\\
0.0608660x^{\frac{3}{5}}-0.0139420x^{-3}& \textrm{if $x \in [R,H)$, $f=-$}\\
0.0116692x^{-\frac{1}{2}} &\textrm{if  $x\in(L,\infty)$, $f=0$}
\end{array} \right.
\end{equation}
Hence when the stock is in the positive regime, the trader buys in the interval $[0.7037890, R]$.  
This is a wider buying price range than the prescription of the standard rule from the support/resistance line method of TA. In the support/resistance line method, the level $R$ is a support level in the positive regime and the trader would receive a buy signal if the price fell to level $R$.

\end{example}
\section{Optimal trading strategies and degrees of relative risk aversion}\label{risk aversion}

In the preceding sections we identified five price levels, namely $B, m, c, b, a$, which together determine the optimal trading strategies. Recall the levels $B, m, c$ are sale thresholds in the positive, negative and zero regime, respectively. The levels $a,b$ are buy thresholds in the positive regime. 
In this section, we explore the relation between these price levels and degrees of relative risk aversion. Numerical methods for solving ODEs are well established and we implement an algorithm to solve the two free boundary problems.

Take the utility function to be a power function of the form $u(x)=x^\gamma$. The price dynamics in the negative and zero regimes are of the form $\m_-(x)=\m_-x$ and $\sigma_-(x)=\sigma_-x$, $\m_0(x)=\m_0 x$ and $\sigma_0(x)=\sigma_0 x$. 
Consider the mean-reverting Vasicek \cite{Vas} model for the dynamics in the positive regime. 
The drift is of the form $\m_+(x)= \xi -\mp x$, and the volatility is of the form $\sigma_+(x)=\sp$ 
for positive constants $\m_+, \sp$ and $\xi$. We have reflection parameters $p_+, p_- $ and the intensity parameter $\l$ governing the probability of entering the zero regime. 

\begin{figure}[h]
\begin{center}
	\includegraphics[scale=0.3]{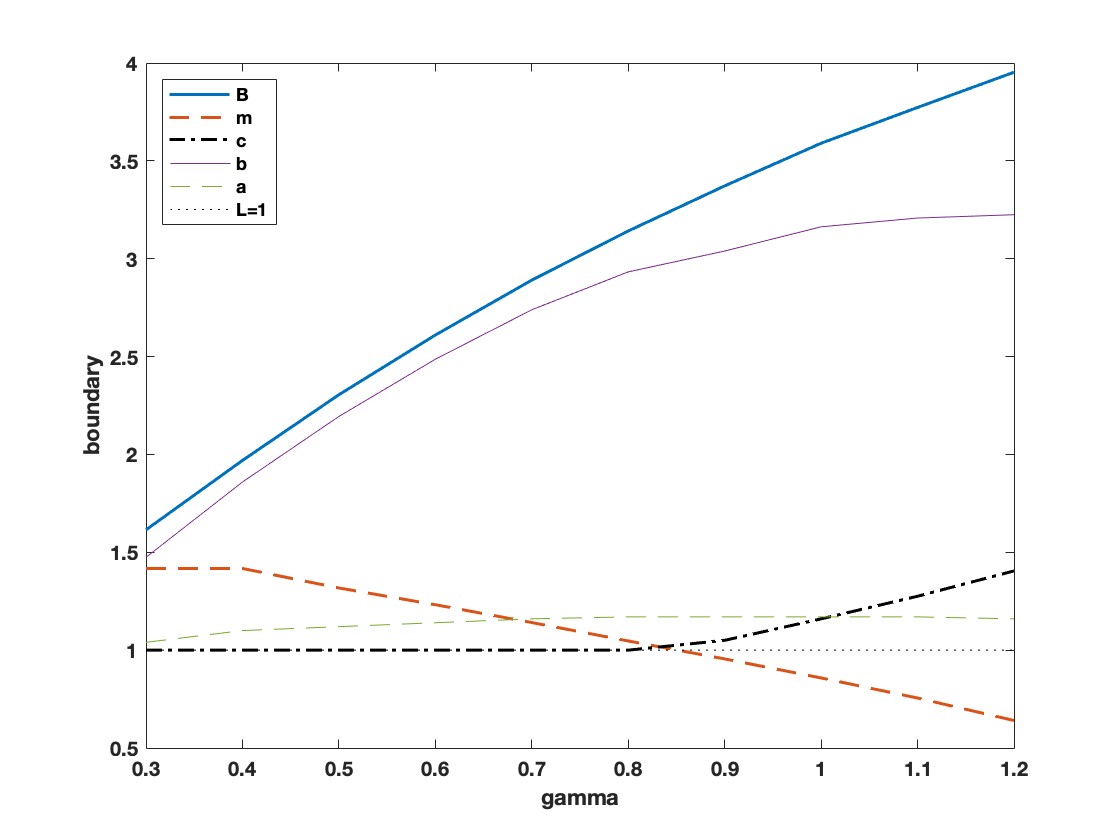}
\end{center}
\caption{Values of boundaries $B, m, c$ and $a,b$ against $\gamma$ for base parameters. 
}
\label{figure:basepars}
\end{figure}

We give a base set of parameters in Table 
\ref{table:pars} and check that the conditions in Assumptions \ref{c} and \ref{suffa} have been met. 
The results are presented in Figure \ref{figure:basepars} where we plot the thresholds $B, m, c, a, b$ against values of $\gamma$ between 0.3 and 1.2. Recall $1-\gamma$ is equal to the relative risk aversion for the power utility, hence, as $\gamma$ increases, the degree of risk aversion decreases.
Note that $c \leq B$ holds, hence we have a unique solution, see Theorems 
\ref{theorem: D shape} and \ref{fbpsoln}.

Firstly, observe the increasing, concave shape of the sale threshold $B$ and buying threshold $b$ in $\gamma$. If the trader is less risk averse, then they are willing to wait for a higher sale price or buy at a higher purchase price, hence the thresholds are increasing. 
A mean-reverting drift would push the stock price down with increasing force as the stock price increases, hence there is a risk associated with waiting for a higher selling boundary $B$ (less chance of getting there) or buying at a higher price $b$ (greater chance of making a loss). 
This makes the trader less willing to increase $B$ or $b$ for each smaller (and eventually negative) degree of relative risk aversion, which results in the concavity. 

Next, observe that the lower sale threshold, $m$, where the trader sells in the negative regime, is decreasing with $\gamma$. A less risk averse trader is willing to wait to sell at a lower price level. 
Note the value of $m$ drops below $L=1$ at around $\gamma \ge 0.83$. 
Moreover, there is a kink for $m$ for $\gamma$ around $0.83$. This is because the boundary condition changes substantially for $m<L$. When 
$m<L$, the trader would continue to hold the stock when the price process transitions from the positive to the negative regime at $L$. 
From Figure \ref{figure:basepars}, this happens when traders are less risk averse and even risk-seeking (i.e. $\gamma>1$), which suggests (at least under our modelling and specifications) waiting for a break-through from the negative to the positive regime is a very risky strategy and should be avoided by more risk-averse traders. 

The threshold $c$, above which the trader sells the stock when in the zero regime, takes the value $c=L=1$ for $\gamma < 0.83$ and is increasing for values of $\gamma$ above this level. 
When risk aversion is stronger ($\gamma < 0.83$), the stock is sold everywhere if it enters the zero regime. The trader does not want to risk waiting to potentially return to the negative regime, because $m>L$ so the stock would be sold immediately upon reaching $L$ anyway. Thus selling immediately upon entering the zero regime in $(L,\infty)$ is optimal.  

For larger values of $\gamma$, when the trader is less risk averse or even risk seeking, then $c$ is above $L=1$. The stock is held in the region $(L,c)$ when in the zero regime, in the hope that the price either rises to $c$ and is sold, or falls to $L$, triggering a transition to the negative regime. 
At larger values of $\gamma$, $m<L$, and the trader waits in the hope that the stock returns to the positive regime, or sells at $m$ if the price falls further. 
The region $(m,c)$ widens beyond the level $\gamma = 0.83$ as $\gamma$ increases - the trader waits for a higher selling level $c$ or a lower selling level $m$, the less risk averse they are.

\begin{table}
	\caption{Base parameter values.}
	\begin{center}
		\begin{tabular}{ c|c|c|c|c|c|c|c|c|c|c|c|c|c } 
			\hline
			$\mp$& $\mm$ & $\sp^2$ & $\sm^2$ & $\xi$ & $r$ & $L$ & $H$ & $\l$  & $R$  & $p_+$ & $p_-$ & $\mo$ & $\so^2$ \\ 
			\hline
			0.1 & 1/30 & 0.1 & 1/30 & 0.7 & 0.1 & 1 & 1.5 & 0.1 & 1.25 & 0.7 & 0.5 & 0.05 & 1/30 \\ 
			\hline
		\end{tabular}
	\end{center}
	\label{table:pars}
\end{table}


\begin{figure}[h]
\begin{center}
	\includegraphics[scale=0.3]{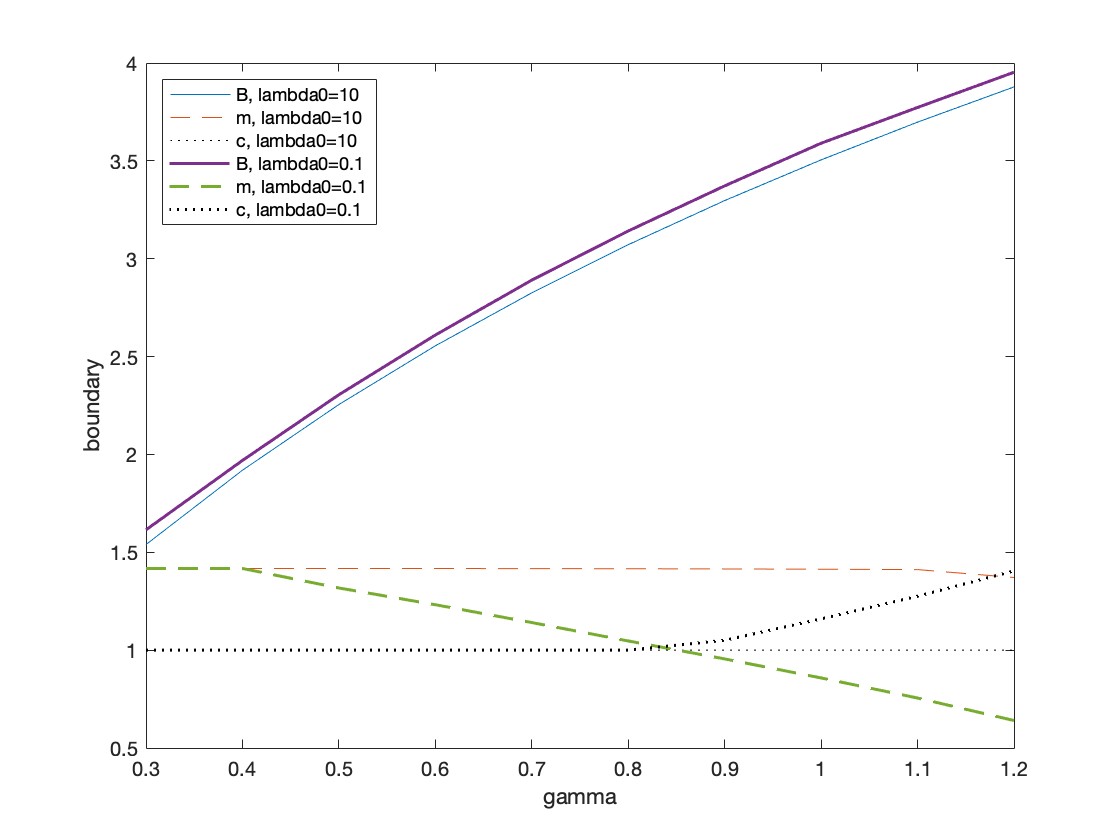}
\end{center}
\caption{Values of boundaries $B, m, c$ against $\gamma$ for two values of the intensity parameter $\l=0.1, 10$. All other parameters are set to their base values.
}
\label{figure:lambdaeffect}
\end{figure}

\begin{figure}[h]
\begin{center}
	\includegraphics[scale=0.3]{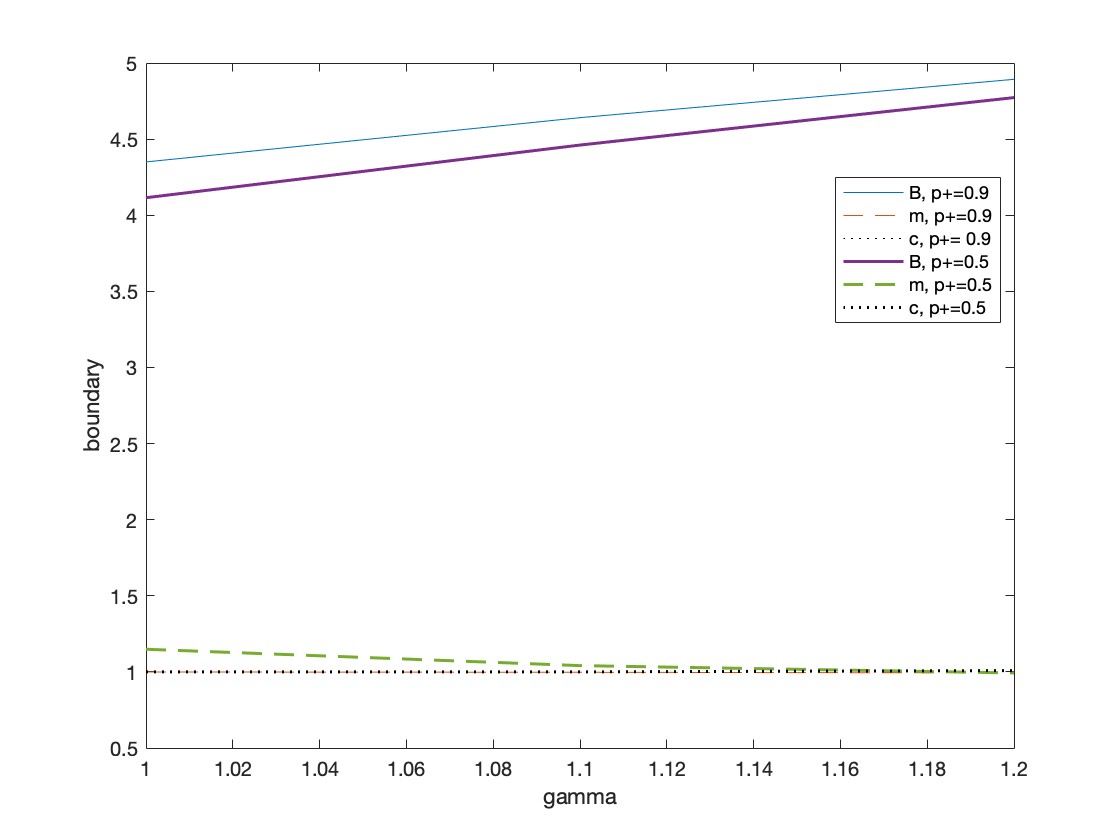}
\end{center}
\caption{Values of boundaries $B, m, c$ against $\gamma$ for two values of reflection parameter $p_+ = 0.5, 0.9$. 
Set $\sp^2 = 1.2$ and all other parameters are set to their base values. }
\label{figure:highvoleffectofpp2}
\end{figure}

\begin{figure}[h]
\begin{center}
	\includegraphics[scale=0.3]{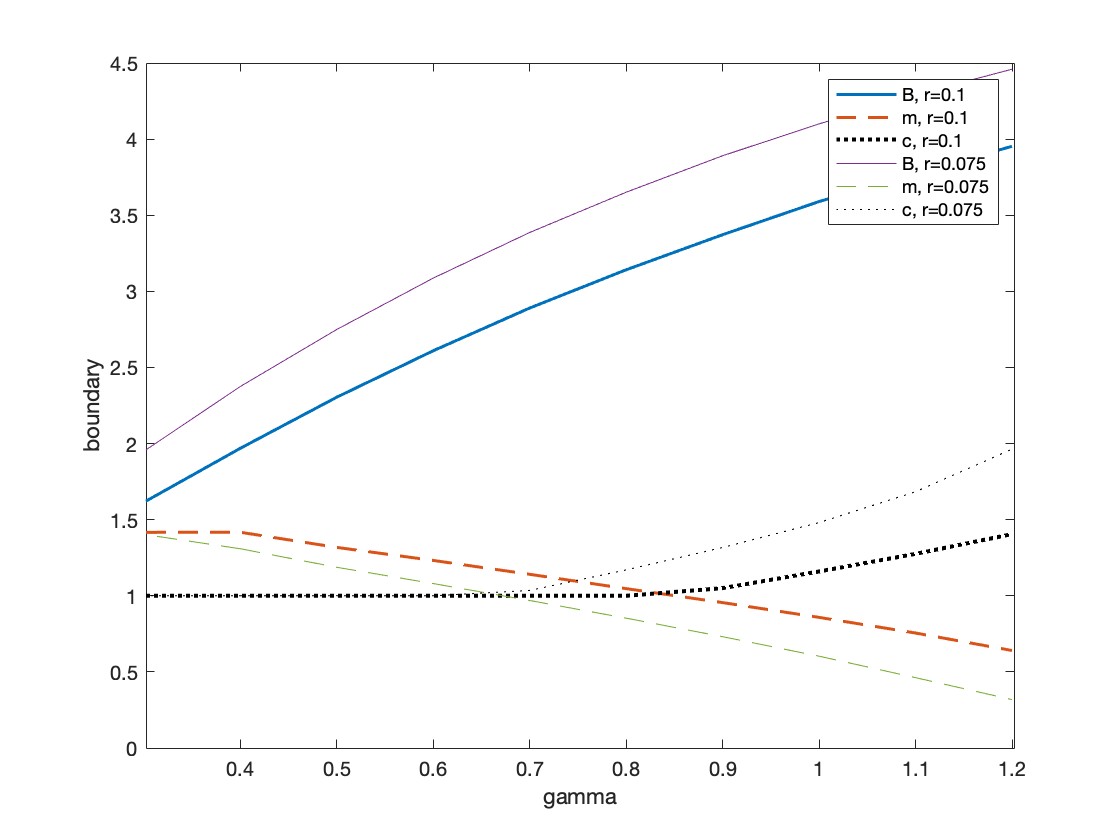}
\end{center}
\caption{Values of boundaries $B, m, c$ against $\gamma$ for two values of interest rate, $r=0.075, 0.10$. All other parameters are set to their base values.  
}
\label{figure:rateeffect}
\end{figure}

In Figure \ref{figure:lambdaeffect} we demonstrate the effect of the intensity parameter $\l$ on the selling boundaries. 
An increased value of $\l$ means a higher probability of entering the zero regime from the positive regime. The effect on the thresholds is as follows. The threshold $B$ at which sales occur in the positive regime is reduced - the agent is less willing to wait for a higher price because of the larger risk of entering the zero regime.  
The lower sale threshold $m$ rises significantly to become almost horizontal at the value 1.41. 
Even at higher values of $\gamma$, when the agent is risk seeking, the agent does not want to wait to sell because of the higher chance of entering the zero regime. In the negative regime, when $\l$ is high, the agent sells at a value around 1.41. 
The threshold $c$, where the agent sells when in the zero regime, also changes with $\l$. 
When the risk of entering the zero regime is higher, the threshold $c=L=1$ even for larger $\gamma$ values when the agent is risk seeking. Hence the stock is sold everywhere in the zero regime, when $\l$ is high.

In Figure \ref{figure:highvoleffectofpp2}, we show the impact of the reflection parameter $p_+$ on selling boundaries. We choose a high volatility in the positive regime, taking $\sp^2 = 1.2$, to enable us to demonstrate a visible movement in thresholds. Then we consider $\gamma \geq 1$ in order to satisfy the constraint on $A$, so the trader is risk seeking. We plot the thresholds $B,m,c$ against values of risk aversion, $\gamma$.

First note that the magnitude of the upper sales thresholds $B$ are greater than in Figure \ref{figure:basepars} due to the higher value of volatility in the positive regime. 
When the reflection parameter in the positive regime is increased, the upper sales threshold $B$, rises for each value of risk aversion. A stronger positive reflection at $R$ causes the trader to wait for a higher price level at which to sell in the positive regime. 
In the negative regime, the sale threshold, $m$, is above $L=1$ when $p_+=0.5$. 
Note that $m$ is higher than in Figure \ref{figure:basepars} due to the higher value of volatility in the positive regime. 

When $p_+$ is increased, the threshold $m$ drops to $L=1$ everywhere. The (risk seeking) trader is willing to wait to risk a lower sale price in the negative regime, because of the slightly better conditions if the price reaches the positive regime ($p_+$ increased). 
Finally, if the price enters the zero regime, then the trader sells everywhere, as the threshold $c=L=1$. 
This is unaffected by the change in $p_+$. 

We demonstrate the effect of changing the interest or discount rate $r$ in Figure 
\ref{figure:rateeffect}. Lowering the rate results in a widening of the selling thresholds - $B$ increases and $m$ decreases. Less discounting being applied means the trader is more willing to wait longer for a higher sale price in the positive regime, or lower sale price in the negative regime.   
We see when the rate is lowered, the sale threshold in the zero regime, $c$, is greater than $L$ for a larger range of risk aversion levels, and the threshold itself increases. 
We note that the region $(m,c)$ is wider for the lowered level of interest rate.

In the negative regime, behaviour depends on the relationship between $m$ and $R$. 
If $m>R$, the stock is sold above the resistance line which is akin to the "sell at high" behaviour in the 
support/resistance TA model. In Figure \ref{figure:basepars}, we see that $m>R$ when the trader is risk averse with $\gamma$ lower than around 0.6. 
However, the model can also generate the situation where $m<R$ and here, the stock is sold at $m$ or below, which is a level below the resistance line. In this case, model behaviour differs from the TA prediction of "sell at high". This occurs when the trader is less risk averse, as seen in Figure \ref{figure:basepars}. 
Another viewpoint is that when the stock price goes above $m$ in the negative regime, it is likely that there will be a break-through. Hence, the current resistance can be thought of as the future support, and the trader now believes the stock is soon going to enter the positive regime. 

\section{Concluding remarks}



We conclude with some caveats. We emphasise that with the introduction of a partially-reflecting boundary, arbitrage opportunities are introduced, i.e. there is no EMM. This makes the model unsuitable for pricing derivatives. Nevertheless, it directly reflects the technical traders' belief about stock price movements, which makes it useful for comparing the output of our model to the standard trading rules from TA. There is widespread belief in the effectiveness of TA, despite incompatibility with the existence of a risk-neutral measure. We believe this is supported by non-equilibrium dynamics and by significant success in practice.

We reiterate that our model cannot reflect all the complexities of path-dependencies of TA but we feel it  is a good compromise between realism and tractability. Although we incorporate path-dependent regime-changes and allow the price to be in any regime in the interval $(L,H)$, we do require fixed levels $L,H$. Whilst the level $R$ captures a support level in the positive regime and a resistance level in the negative regime, the level itself is fixed rather than being dynamically updated during a trading episode.   
There is much scope for future research to address these limitations in a richer model.

\appendix
\section{Proofs and additional results}

\subsection{Proof of Theorem \ref{fund}}\label{fundproof}

\begin{proof}[Proof of Theorem \ref{fund}]

We first construct a process $(S^\pm,F^\pm)$ which has the same dynamics as $(S,F)$ in the positive and negative regimes and does not transition to the 0 regime (this corresponds to the case where $\l=0$). 

We construct the resolvent of $S^\pm$ via the obvious iteration scheme and then deduce the other properties from those of the resolvent. For ease of notation, we denote the unique element of $(+,-)\setminus\{f\}$ by $-f$ and temporarily relabel $L$ and $H$ by $F^-$, $F^+$ respectively.

Denote the resolvent of $S^f$ (killed at $F^{-f}$) by $R^f_\la$ and define $R^{f,n}_\la$ (acting on $C_b(\R\cup \{\partial\},\R)$) inductively by
\begin{equation}\label{iter}
R^{f,1}_\la=R^{f}_\la;\;\; R^{f,n+1}_\la g(x)=R^f_\la g(x)+(1-\la R_\la^f1(x))R_\la^{-f,n}g(F^{-f}),
\end{equation}
where $1(x):= I_{(x\neq \partial)}$.
It should be clear that $R^{f,n}$ corresponds to a process which looks like the desired $S^\pm$ but dies on the $n$th regime switch.

Now we take limits in $n$ in equation (\ref{iter})
$$
\tilde R^f:= \lim_{n\rightarrow\infty}R^{f,n}.
$$
The limit is guaranteed to exist since $0\leq (1-\la R_\la^{f}1(F^{-f}))<1$.

Now define $\bar R$ by
$$
\bar R g(\cdot,f)=\tilde R^f g(\cdot,f)\text{ for }f=\pm.
$$
It is easy to check from this definition that $\bar R$ 
is a contraction resolvent on $C_b(E,\R)$ (see \cite{Will} III.4) and is the unique contraction resolvent $T$ on $C_b(E,\R)$ (bounded continuous functions from $E$ to $\R$) satisfying 
\begin{equation}\label{rdef}
T_\la g(x,f)=R_\la^f g(x,f)+ (1-\la R_\la^f 1(x))T_\la g(F^f,-f).
\end{equation}
It follows from the definition that
\begin{equation}\label{res}
\bar R_\la1(F^f)=R^f_\la 1(F^f)+\bigl(1-\la R^f_\la 1(F^f)\bigr)\biggl(R^{-f}_\la 1(F^{-f})+(1-\la R^{-f}_\la 1(F^{-f}))\bar R_\la1(F^f)\biggr),
\end{equation}
and substituting into (\ref{rdef}) we see that $\la \bar R_\la 1(F^f)=1$.
 It follows from  (\ref{rdef}) that
$$
\la\bar R 1_E(x,f)=1\text{ for all }(x,f)\in E,
$$
and so $\bar R$ is conservative.

To show that $\bar R$ is the resolvent of a conservative transition semigroup $(P_t)_{t\geq 0}$, it remains to show that $\la\bar R$ is positive and a contraction on $C_b(E,\R)$ (equipped with the sup-norm, $||\cdot||_{\infty}$).
Positivity follows immediately from the positivity of $R^f$ and the iteration (\ref{iter}).
Contractivity follows from the contractivity of $R^f$ by induction and the fact that $||R^f g||_{\infty}\leq ||g||_{\infty}||R^f1||_{\infty}$ (which follows from positivity).

Using the identity
$$
\G g=\lim_{\l\rightarrow \infty}\la (\la R_\la-I)g,
$$
where $R$ is a resolvent corresponding to the infinitesimal generator $\G$ and $g\in\D(\G)$ (see (4.12) p111 in \cite{Will}), it is easy to deduce from (\ref{res}) that the infinitesimal generator of the semigroup is $\L$ and hence that $(S^{\pm},F^{\pm})$ satisfies a suitably simplified version of (\ref{defS}) and \ref{defF}).

To show that $(P_t)$ is Feller, it remains (see p 166 of \cite{EK}) to show that the transition semigroup is strongly continuous, but this follows from the strong continuity of the semigroups for $S^f$, $f=\pm$.
Theorem 2.7 of Ch.4 of \cite{EK} now tells us that $(S^\pm,f)$ is strong Markov.

The c\`adl\`ag property follows from the fact that $S^f$ is continuous up to its death time (the first hitting time of $F^{-f}$), and so the time for $F_t$ to jump is always strictly positive unless $(S^\pm_{t-},F_{t-})=(F^{-f},f)$, while these times cannot cluster since $(P_t)$ is conservative.
Finally, the desired regularity follows from that of $S^f$, $f=\pm$. 

The construction of (the law of) $(S,F)$ from the resolvents of $(S^\pm,F^\pm)$, killed at variable rate 
$\hat\l(\cdot, f)=1_{(f=+)}\l$, 
and $S^0$, killed on hitting $L$, and the proof of the requisite properties of $(S,F)$ follows in much the same way.

\end{proof}

\subsection{Additional lemmas}\label{addl}

\begin{lemma}\label{point}
	Let
	\begin{displaymath}
	\gg_x=\{\omega:\; \forall t>0\; \exists s\in(0,t)\text{ with } S_s(\omega) >x \}.
	\end{displaymath}
	Analogously, we define 
	\begin{displaymath}
	\ll_x=\{\omega:\;   \forall t>0\; \exists s\in(0,t)\text{ with } S_s(\omega) <x\}.
	\end{displaymath}
	Then, $\P_{x,f}(\gg_x)=1$ and $\P_{x,f}(\ll_x)=1$, for each $(x,f) \in E\setminus(0,-)$.
\end{lemma}

\begin{proof}
	The strategy is similar to the proof of (6) in Chapter 2 of Freedman \cite{Fre}.\\
	By Blumenthal's 0-1 Law, $\P_{x,f}(\gg_x)$ is either $1$ or $0$. Suppose it is 0. We set $\t:=\inf\{t \ge 0:\;   S_t >x\}$. Then, $\P_{x,f}(\t=0)=0$ since $\{\t=0\}= \gg_x$. Moreover, it is not hard to check $\{\t=\infty\}=\{S_t \le x \text{ for all } t\}$, $S_\t=x$ on $\{\t<\infty\}$, and $\t=0$ on $\{\t < \infty\}$ given starting position $S_\t$. Therefore, by the strong Markov property, 
	\begin{equation}
	\begin{split}
	\P_{x,f}(\tau < \infty)&=\P_{x,f}(\tau < \infty, \t=0 \text{ given starting position } S_\t)\\
	&=\P_{x,f}(\t=0 \text{ given starting position } S_\t \ |\ \tau < \infty)\P_{x,f}(\tau < \infty)\\
	&=\P_{x,f}(\t=0)\P_{x,f}(\tau < \infty)=0.
	\end{split}
	\end{equation}
	Thus, $\P_{x,f}(\tau = \infty)=1$, and hence $\P_{x,f}(S_t \le x \text{ for all } t)=1$, which contradicts the regularity of $S$. Therefore, $\P_{x,f}(\gg_x)=1$. The proof for the event $\ll_x$ is symmetric. 
\end{proof}

\begin{lemma}\label{appendix: equivalence}
	$\V_p$ defined by (\ref{prob2}) has an equivalent formulation as 
	\begin{equation}
		\V_p(x,f):=\sup_{\tau}\E^{(x,f)}[e^{-r\tau}\{\V(S_\tau,F_\t)-u(S_\t)\}],
	\end{equation}
	where $\V$ is the solution to the seller's problem.
\end{lemma}

\begin{proof}
  Note that $e^{-r\tau}\V(S_\tau,F_\t) = \esssup_{\t_0 \ge \t} \E[e^{-r\tau_0}u(S_{\t_0}) | \F_{\t}]$.  Let $Z^\t = \E[e^{-r\t}u(S_{\t})|\F_{\t_1}]$ and $Z^* := \esssup_{\t \ge \t_1} \E[e^{-r\tau}u(S_{\t}) | \F_{\t_1}]]$. It is sufficient to show that 
  \begin{equation}
  	\sup_{\t_1 \le \t_2 } \E[Z^{\t_2}] \ge \sup_{\t_1}\E[Z^*].
  \end{equation}
 For arbitrary stopping times $\t \ge \t_1$ and $\stop \ge \t_1$, define the stopping time $\t^0$ by
 \begin{equation}
 	\t^0 = \t \1_{Z^\t \ge Z^\stop} + \stop \1_{Z^\t < Z^\stop}. 
 \end{equation}
 Hence, $Z^{\t^0} =\E[e^{-r\t}u(S_{\t})\1_{Z^\t \ge Z^\stop} +e^{-r\stop}u(S_{\stop}) \stop \1_{Z^\t < Z^\stop} |\F_{\t_1}]  \ge \max \{Z^\t, Z^\stop\}$. Thus, there is an increasing sequence of stopping times $\stop_n$ such that $Z^{\stop_n}$ increases to $Z^*$. Moreover, since
  \begin{equation}
 \E[|Z^*|] \le \E [\esssup_{\t \ge \t_1} \E[|e^{-r\tau}u(S_{\t})| | \F_{\t_1}]]] \le \E[\sup_{t}|e^{-rt}u(S_{t})|] < \infty,
 \end{equation}
 by the Monotone Convergence Theorem, we conclude
 \begin{equation}
 \sup_{\t_1}\E[Z^*] =  \sup_{\t_1} \lim_{n \rightarrow \infty} \E[Z^{\stop_n}] \le  \sup_{\t_1}  \sup_{\t_2 \ge \t_1} \E[Z^{\t_2}] = 	\sup_{\t_1 \le \t_2 } \E[Z^{\t_2}]. 
 \end{equation}
\end{proof}

\subsection{Proof of Theorem \ref{condition}}\label{addl}
We shall appeal to the following result which is mentioned in \cite{J} but for which no proof was given.
\begin{thm}\label{bound}
Suppose that $Z$ is a continuous, positive process adapted to $(\Omega,\F,(\F_t),\P)$. Define $S$ to be the running maximum of $Z$ so that
$$
S_t:=\sup_{s\leq t} Z_s.
$$
Further suppose  that there is a $p>1$ and a sequence of stopping times $T_n\uparrow\infty$ a.s.such that
\begin{equation}\label{max}
\lim_n\sup_{\tau\leq T_n}\E[Z^p_\tau]<\infty,
\end{equation}
then
$$
\E[S_\infty]<\infty.
$$
\end{thm}
\begin{proof}
For each $x\in (0,\infty)$ define
$$
\tau^n_x=\min(\inf\{t\leq T_n:\;\; Z_t\geq x\},T_n).
$$
Clearly $(S_{T_n}\geq x)=(Z_{\tau^n_x}\geq x)$.

Take $C:\;\sup_{\tau\leq T_n}\E[Z^p_\tau]\leq C$ for all $n$ (which we can, since the limit in (\ref{max}) is a monotone one) then, by Markov's inequality,
\begin{equation}\label{max2}
\P(S_{T_n}\geq x)=\P(Z_{\tau^n_x}\geq x)\leq \frac{\E[Z^p_{\tau^n_x}]}{x^p}\leq \frac{C}{x^p}.
\end{equation}
Now, using the standard result that, if $X$ is a non-negative random variable,
$$
\E[X]=\int_0^\infty \P(X\geq x)dx,
$$
we deduce from (\ref{max2} that
$$
\E[S_{T_n}]\leq 1+\int_1^\infty \frac{C}{x^p}dx=1+\frac{C}{p-1}.
$$
Since $S$ is an increasing process, we see, by the Monotone Convergence Theorem, that
$$
\E[S_{\infty}]\leq 1+\frac{C}{p-1}.
$$

\end{proof}
\begin{remark}
Continuity is not actually required in Theorem \ref{bound}. The argument only needs small modifications if $Z$ is just predictable, on appealing to the Predictable Section Theorem.
\end{remark}
\begin{proof}[Proof of Theorem \ref{condition}]
Take $U$ to be a positive, increasing $C^2$ function with
$$
U(x)\begin{cases}=u(R+2)&\text{ for }x\leq R+1\\
	\geq u(x)&\text{ for }x\in (R+1,R+3)\\
	=u(x)&\text{ for }x\geq R+3
\end{cases}
$$
We define  a positive, continuous  process $Z$ by
$$
Z_t=e^{-rt}U(S_t),
$$
and set $Q=Z^p$. Then, from Ito's Lemma we see that
\begin{equation}\label{diff1}
dQ_t=e^{-prt}pU(S_t)^{p-1}\biggl( (\mL^{F_t}-r)U(S_t)+\half (p-1)\frac{(\sigma_{F_t}U'(S_t))^2}{U(S_t)}\biggr)dt +dN_t,
\end{equation}
where $N$ is a local martingale.
Since $U$ is $C^2$ by assumption, denoting $\{(x,f)\in E^-\times\{-\}\cup(L,M]\times\{+,0\}\}$ by $E^M$, 
$$
\sup_{E^M}\biggl(( \mL^{F_t}-r)U(S_t)+\half (p-1)\frac{(\sigma_{F_t}U'(S_t))^2}{U(S_t)}\biggr)=\kappa_M<\infty,
$$
for any $M\geq \max(A,R+3)$. Now, thanks to (Assumption \ref{suffa}), we may take $M$ such that 
$\sup_{x\geq M}\max\biggl[\biggl(\frac{\sigma_+(x)U'(x)}{U(x)}\biggr)^2,\biggl(\frac{\sigma_0(x)U'(x)}{U(x)}\biggr)^2\biggr]=D$ and $\max\bigl[\bigl(\mL^+U(x)-rU(x)\bigr),\bigl(\mL^0U(x)-rU(x)\bigr)\bigr]\leq -\epsilon U(x)$ for all ${x\geq M}$.

Then denoting $\sup_{E^M}U(x)$ by $d_M$,
\begin{equation}\label{diff2}
dQ_t\leq dN_t+e^{-prt}p\biggl(d_M^{p-1}\kappa_M-U(S_t)^{p}[\epsilon-\half(p-1)D]1_{(S_t\geq M)}\biggr)dt.
\end{equation}
Setting $p=1+\frac{\epsilon}{D}$ in (\ref{diff2}, and taking  a localising sequence $T_n$ for the local martingale $N$
we see that $E[Q_\tau]\leq \frac{d_M^{p-1}\kappa_M}{r}$, for any stopping time $\tau$ bounded by $T_n$. Since the bound is independent of $n$, we conclude that $Q$ satisfies (\ref{max}) and thus, by Theorem \ref{bound}, condition \ref{sup} of  Assumption \ref{assumption: utility} holds.

To prove condition \ref{limit} of  Assumption \ref{assumption: utility}, note that, since  $e^{-rt}\V(S_t,F_t)$ is a non-negative, class D supermartingale it has a (non-negative) limit $X$ almost surely and in $L^1$. 
Now define $$
Y:= \limsup e^{-rt}u(S_t).
$$
If $\E^{(x,f)}[ X]=0$ for some, and then, by irreducibility, for all $(x,f)\in E\setminus\{(0,-)\}$ then, since $\V$ dominates $u$, we must have $Y=0$ a.s. and in $L^1$, so we now suppose that $\E^{(x,f)}[ X]>0$ for some, and then, by irreducibility, for all $(x,f)\in E\setminus\{(0,-)\}$. Note that, on the event $X>0$ we have $\limsup u(S_t)=\limsup S_t=\infty$.
\begin{itemize}
\item[1]{\it The case $\l=0$:} Define $\low=\max(A,R+3,H)$ and suppose that $x>\low$.
Since $Y=Y1_{(\t^+_{\low}=\infty)}+Y1_{(\t^+_{\low}<\infty)}$, we can apply the strong Markov property to deduce that
$$
\E^{(x,+)}[Y]=\E[Y1_{(\t^+_{\low}=\infty)}]+\E^{(x,+)}[e^{-r\t^+_x}]\E^{(x,+)}[Y]\Rightarrow \E^{(x,+)}[Y]=\frac{\E[Y1_{(\t^+_{\low}=\infty)}]}{1-\E^{(x,+)}[e^{-r\t^+_x}]}.
$$
Now, since $e^{-r{t\wedge \tau^+_\low}}u(S_{t\wedge \tau^+_\low})$ is a class D supermartingale 
(since $\low\geq A$) we see that $Z_{t\wedge \tau^+_\low}$ is a positive class D supermartingale (since $\low\geq R+3$) so
converges almost surely and in $L^1$ to $Y1_{(\t^+_{\low}=\infty)}+e^{-r(\t^+_{\low})}u(\low)1_{(\t^+_{\low}<\infty)}$ and so 
$$
\lim_t\E^{(x,+)}[Z_{t\wedge \tau^+_\low}]= \E^{(x,+)}[Y1_{(\t^+_{\low}=\infty)}+e^{-r(\t^+_{\low})}u(\b)1_{(\t^+_{\low}<\infty)}]\geq \E^{(x,+)}[Y1_{(\t^+_{\low}=\infty)}].
$$
Now,
\begin{equation}\label{cont}
\E^{(x,+)}[Z_{t\wedge \tau^+_\low}]=u(x)+\E^{(x,+)}[\int_0^{t\wedge \tau^+_\low}e^{-rs}(L^+u-ru)(S_s)ds]\leq
u(x)-\e\E^{(x,+)}[\int_0^{t\wedge \tau^+_\low}Z_s1_{(S_s\geq \Delta)}ds],
\end{equation}
by Assumption \ref{suffa}. But on the event $\t^+_\low=\infty$, $\lim_t Z_t=Y$, and on the event $(Y1_{\t^+_\low=\infty}>0)$ we must have $\lim_tS_t=\infty$ a.s., so if $\P(Y1_{\t^+_\low=\infty}>0)>0$, the RHS of (\ref{cont}) converges to $-\infty$, which is a contradiction, since $Z\geq 0$.

\item[2]{\it The case $\l>0$:} this is very similar except we consider the 0-regime. In this case we decompose $Y$ as $Y=Y1_{(\t^0_{\low}=\infty)}+Y1_{(\t^0_{\low}<\infty)}$.
We apply the strong Markov property to deduce that
$$
\E^{(x,0)}[Y]=\E[Y1_{(\t^0_{\low}=\infty)}]+\E^{x,0}[e^{-r\t^0_x}]\E^{x,0}[Y]\Rightarrow \E^{x,0}[Y]=\frac{\E[Y1_{(\t^0_{\low}=\infty)}]}{1-\E^{x,0}[e^{-r\t^0_x}]}.
$$
Now if $\limsup Z_t>0$, $S$ must visit the 0-regime at arbitrarily large times and we can deduce a contradiction in exactly the same way as in Case 1, by considering $Z_{t\wedge \t^0_\low}$ from starting position $(x,0)$ which is then  a class D supermartingale since $\low\geq L$.
\end{itemize}
\end{proof}

\end{document}